\documentclass[lettersize,journal]{IEEEtran}
\usepackage{amsmath,amsfonts}
\usepackage{array}
\usepackage{hyperref}
\usepackage[caption=false,font=footnotesize,labelfont=rm,textfont=rm]{subfig}
\usepackage{bbding}
\usepackage{textcomp}
\usepackage{stfloats}
\usepackage{url}
\usepackage{verbatim}
\usepackage{graphicx}
\usepackage[linesnumbered,ruled,vlined]{algorithm2e}
\usepackage{cite}
\usepackage{multirow}
\usepackage{makecell}
\usepackage{booktabs} 
\usepackage{threeparttable}
\def\BibTeX{{\rm B\kern-.05em{\sc i\kern-.025em b}\kern-.08em
    T\kern-.1667em\lower.7ex\hbox{E}\kern-.125emX}}
\begin{document}

\title{MER-CLIP: AU-Guided Vision-Language Alignment for Micro-Expression Recognition}

\author{
        Shifeng~Liu,
        Xinglong~Mao,
        Sirui~Zhao*,
        Peiming~Li,
        Tong~Xu,~\IEEEmembership{Member,~IEEE,}\\
        and~Enhong~Chen*,~\IEEEmembership{Fellow,~IEEE}
\IEEEcompsocitemizethanks{\IEEEcompsocthanksitem Shifeng Liu and Xinglong Mao are with School of Artificial Intelligence and Data Science, University of Science and Technology of China, Hefei, Anhui 230027, China. E-mail: \{lsf0619, maoxl\}@mail.ustc.edu.cn
\IEEEcompsocthanksitem Sirui Zhao, Tong Xu, and Enhong Chen are with School of Computer Science and Technology, University of Science and Technology of China, Hefei, Anhui 230027, China. E-mail: \{siruit, tongxu, cheneh\}@ustc.edu.cn
\IEEEcompsocthanksitem Peiming Li is with School of Computer Science and Technology, Shandong University, Qingdao, Shandong 266237, China.\protect
}

\thanks{(\textit{* Corresponding authors: Enhong Chen and Sirui Zhao}).}
\thanks{This work has been submitted to the IEEE for possible publication. Copyright may be transferred without notice, after which this version may no longer be accessible.}
\thanks{This work was fully supported by the National Natural Science Foundation of China (No.62406264, 61727809, 62072423).} 

\thanks{Manuscript received December xx, xx; revised xx xx, xx.}}

\markboth{Journal of \LaTeX\ Class Files,~Vol.~14, No.~8, August~2021}%
{Shell \MakeLowercase{\textit{et al.}}: A Sample Article Using IEEEtran.cls for IEEE Journals}


\maketitle

\begin{abstract}
As a critical psychological stress response, micro-expressions (MEs) are fleeting and subtle facial movements revealing genuine emotions. Automatic ME recognition (MER) holds valuable applications in fields such as criminal investigation and psychological diagnosis. The Facial Action Coding System (FACS) encodes expressions by identifying activations of specific facial action units (AUs), serving as a key reference for ME analysis. However, current MER methods typically limit AU utilization to defining regions of interest (ROIs) or relying on specific prior knowledge, often resulting in limited performance and poor generalization. To address this, we integrate the CLIP model’s powerful cross-modal semantic alignment capability into MER and propose a novel approach namely MER-CLIP. Specifically, we convert AU labels into detailed textual descriptions of facial muscle movements, guiding fine-grained spatiotemporal ME learning by aligning visual dynamics and textual AU-based representations. Additionally, we introduce an Emotion Inference Module to capture the nuanced relationships between ME patterns and emotions with higher-level semantic understanding. To mitigate overfitting caused by the scarcity of ME data, we put forward LocalStaticFaceMix, an effective data augmentation strategy blending facial images to enhance facial diversity while preserving critical ME features. Finally, comprehensive experiments on four benchmark ME datasets confirm the superiority of MER-CLIP. Notably, UF1 scores on CAS(ME)$^3$ reach 0.7832, 0.6544, and 0.4997 for 3-, 4-, and 7-class classification tasks, significantly outperforming previous methods.
\end{abstract}


\section{Introduction}
\begin{figure}[t]
\centering
\subfloat[MaE]{\includegraphics[width=0.9in]{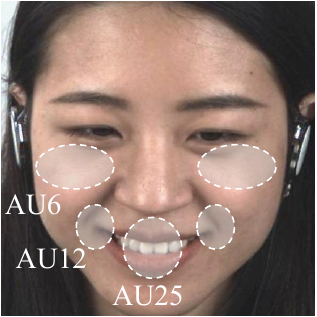}%
\label{fig_ma}}
\hfil
\subfloat[ME]{\includegraphics[width=0.9in]{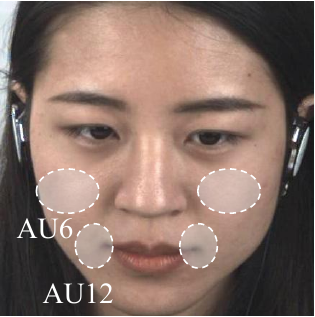}%
\label{fig_mi}}
\caption{Examples of a MaE (a) and a ME (b) from the same person in the DFME database \cite{zhao2023dfme}, both reflecting the `happiness' emotional state. To facilitate observation, we have annotated their activated action units related to specific facial muscle movements based on FACS. The MaE includes a combination of Cheek Raiser (AU6), Lip Corner Puller (AU12), and Lips Part (AU25). The ME only includes slight AU6 and AU12.}
\label{fig_ma_mi}
\vspace{-0.5cm}
\end{figure}

\IEEEPARstart{M}{icro}-expressions (MEs) are a type of facial expression characterized by their brief duration (typically less than 0.5s) and subtle intensity \cite{ekman1969nonverbal,yan2013fast}. As illustrated in Figure~\ref{fig_ma_mi}, unlike macro-expressions (MaEs), which involve more prominent and widespread facial muscle movements, MEs are fleeting and often localized to specific facial regions, making them barely noticeable to untrained eyes.
Moreover, from a psychological perspective, MEs are regarded as important indicators of psychological stress reactions \cite{rinn1984neuropsychology,ben2021video}. They are involuntary and cannot be faked, making them a reliable reflection of a person's true emotions \cite{ekman1969nonverbal}. Therefore, the task of automatic ME recognition (MER) has attracted much attention due to its significant application value in various fields such as criminal investigation, financial decision-making, psychological diagnosis, and education \cite{ekman2009telling,weinberger2010airport,hunter2020emotional,haselhuhn2014negotiating}. Specifically, a MER model is tasked with classifying a single image or a trimmed video sequence into a discrete emotion category. However, due to the subtle nature of ME movements, as well as the challenges posed by significant intra-class variation and inter-class similarity, accurately capturing and distinguishing these fleeting facial muscle movements, meanwhile modeling their relationship to emotions is a crucial but difficult task.

Building on the anatomical basis of facial movements, Ekman and Friesen developed the Facial Action Coding System (FACS) in the 1970s \cite{ekman1978facial}. This system encodes any expression by identifying the activation of specific facial action units (AUs) related to individual muscles or muscle groups. Over the past few decades, FACS has been widely used in various aspects of facial expression analysis. Since AUs support comprehensive coding of subtle movements in localized facial regions, they are crucial for capturing the minute changes in MEs. Consequently, to facilitate ME analysis, most ME datasets provide manually annotated AU labels according to FACS \cite{Davison2018SAMMAS,Yan2014CASMEIA,ben2021video,li20224dme,li2022cas,zhao2023dfme}. Quite a few MER approaches have attempted to incorporate AUs to aid in learning ME motion features \cite{polikovsky2009facial,pfister2011recognising,liu2015main,liu2018sparse,su2021key,wang2024jgulf,Zhao2021AT3,aouayeb2021micro,thuseethan2023deep3dcann,li2020joint,zhai2023feature}. These methods generally utilize commonly observed AUs to define several facial regions of interest (ROIs), followed by specifically extracting features from these regions for ME motion representation learning. However, they typically rely on strong priors and require complex preprocessing, while often overlooking the global information across the entire face. Additionally, some researchers \cite{lo2020mergcnmr,Lei2021MicroexpressionRB,kumar2021microexpressioncb,zhang2023adaptive,kumar2024uncovering} have integrated AUs into graph structures and employed Graph Neural Networks (GNNs) for MER, but the complexity of graph networks often leads to overfitting and poor generalization.

\IEEEpubidadjcol
In fact, AUs not only serve as the encoding of facial muscle movements but also carry clear semantic information and can be described textually in terms of their meanings (e.g., \textit{raising eyebrows}, \textit{tightening lips}) \cite{ekman1978facial}, which introduces a new perspective for MER. With this in mind, we attempt to incorporate the powerful multimodal framework, Contrastive Language-Image Pretraining (CLIP) \cite{radford2021learning} model into MER. By utilizing large-scale paired image-text data, CLIP aligns visual and textual representations through a contrastive learning objective. Drawing inspiration from this approach, we harness the detailed description of localized facial movements provided by AUs to aid in capturing subtle MEs.

However, directly applying CLIP to MER presents several challenges.
Firstly, extracting spatiotemporal representations is crucial for understanding the emotional significance of MEs. Nevertheless, the original CLIP vision encoder is designed for static images \cite{radford2021learning}, which lacks the ability to encode temporal features in videos. Secondly, emotion features for MER require a higher level of semantic understanding than AU-based motion features. Instead of a clear correspondence between AUs and emotions, different AU combinations can correspond to the same emotion category, while similar AU movements can be associated with different emotions \cite{dong2021brief,ben2021video,zhao2023dfme}. Therefore, it is necessary to further bridge the semantic gap between the motion features described by AUs and emotion features for MER. 
Finally, the success of CLIP is largely attributed to the availability of large and diverse training data \cite{radford2021learning}, while ME datasets are typically small in scale. Unfortunately, advanced effective visual data augmentation techniques like MixUp \cite{zhang2017mixup}, CutOut \cite{devries2017improved}, and CutMix \cite{yun2019cutmix} can cause varying degrees of disruption to the subtle and fragmented ME movements. Thus, there is an urgent need for an effective data augmentation strategy during the MER training.

\begin{figure*}[t] 
    \centering
    \includegraphics[scale=0.49]{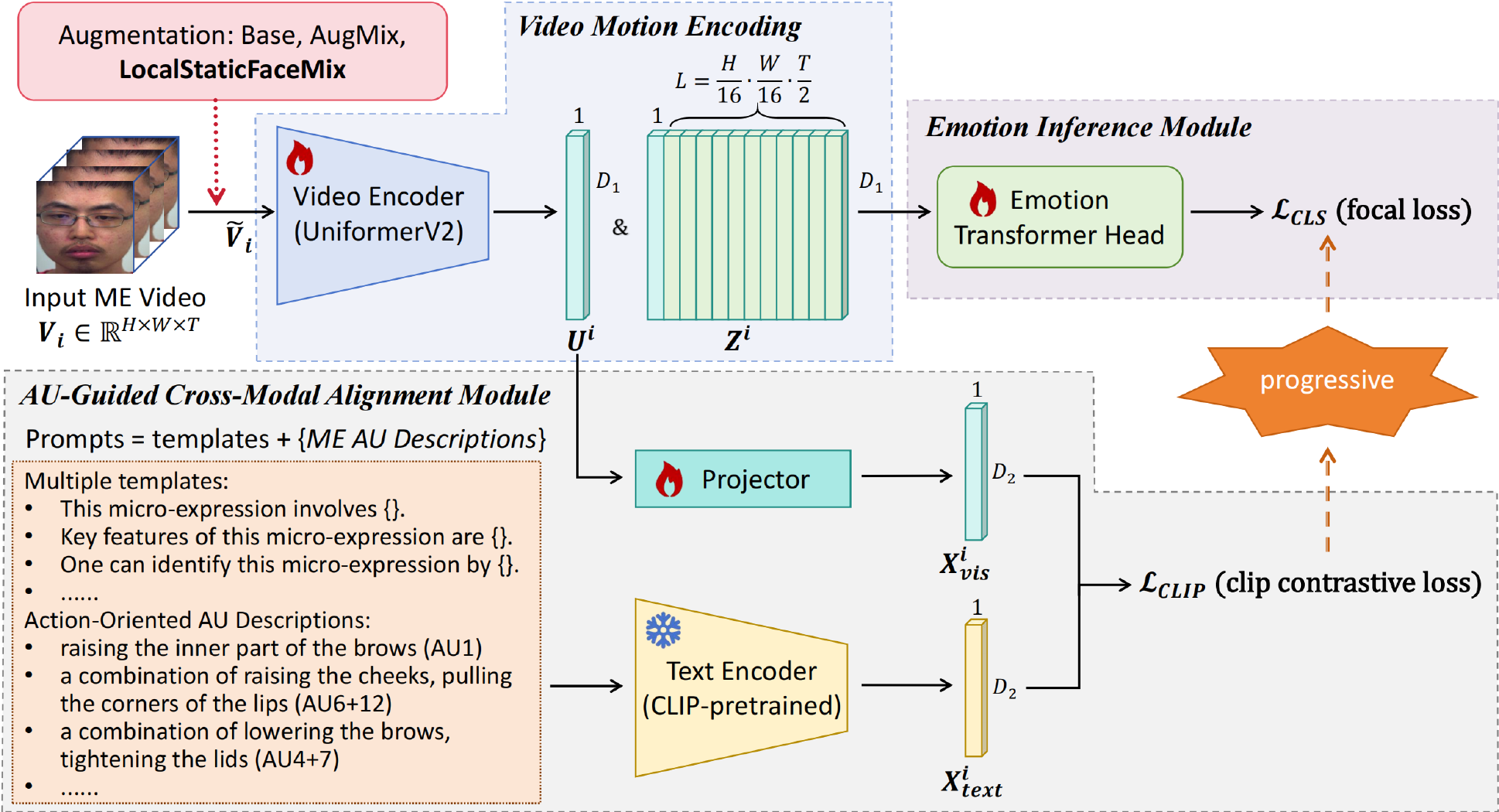}
    \caption{We illustrate the detailed architecture of MER-CLIP. Before being fed into the video encoder, the input video undergoes a series of augmentations, including the basic transformations such as color jitter and flipping, Augmix, and our proposed LocalStaticFaceMix. The augmented video is then passed through the Video Motion Encoding to extract the spatiotemporal ME movement features. These features are subsequently sent into two main components: the AU-Guided Cross-Modal Alignment Module and the Emotion Inference Module to perform fine-grained cross-modal alignment and emotion learning.}
    \label{fig:framework}
    \vspace{-0.4cm}
\end{figure*}

To address aforementioned challenges, we propose \textbf{MER-CLIP}, a CLIP-based model tailored for MER with dedicated architectural enhancements. The detailed framework is shown in Figure \ref{fig:framework}. Specifically, to enhance spatiotemporal feature learning, we replace the original CLIP vision encoder with UniformerV2 \cite{li2022uniformerv2} for ME video motion encoding. Additionally, we generate fine-grained textual descriptions of facial muscle movements for each ME based on available AU labels and feed them into the pre-trained CLIP text encoder. By aligning the visual and textual features in a shared embedding space, this module enhances the vision encoder's ability to capture subtle ME dynamics, which is referred to as the AU-Guided Cross-Modal Alignment Module. In the Emotion Inference Module, we introduce a lightweight transformer head instead of a common linear classification head after the vision encoder, enabling abstract emotion feature learning and final MER. Additionally, we develop a customized data augmentation technique for ME movements, called LocalStaticFaceMix. This operation randomly and partially blends each frame of an ME video sequence with the onset frame from another sequence, which can be seen as a static facial image. In this way, we not only enhance facial appearance diversity, mitigating overfitting on small-scale facial datasets, but also minimize disruption of the original ME motion patterns.

Conclusively, our contributions can be summarized as:
\begin{itemize}
\item We introduce MER-CLIP, an AU-Guided visual-language alignment method for MER. By converting AU labels into detailed textual descriptions of ME movements, we establish fine-grained cross-modal semantic alignment between visual dynamics and textual AU-based interpretations of MEs.
\item To bridge the semantic gap between motion features and higher-level emotional representations, we introduce the Emotion Inference Module, which refines the model’s comprehension of emotional semantics and significantly improves MER performance. 
\item We design a specialized data augmentation technique for ME analysis, namely LocalStaticFaceMix, which effectively reduces facial identity interference and mitigates overfitting on small-scale ME datasets while preserving subtle ME motion features.
\item Extensive experiments on four benchmark ME datasets demonstrate that MER-CLIP surpasses state-of-the-art methods across multiple classification tasks. Comprehensive ablation studies and visualization analyses further validate the effectiveness of our proposed modules.
\end{itemize}

The rest of this paper is organized as follows. In section \ref{sec_related_work}, we first summarize currently existing AU-assisted MER methods, then we introduce related works on CLIP and ME data augmentation. In Section \ref{sec_method}, we provide a detailed explanation of the key components of MER-CLIP. Then the comprehensive experiments is developed and discussed in Section \ref{sec_experiment}. Finally, research conclusions and future work are addressed in Section \ref{sec_conclusion}.

\vspace{-0.4cm}
\section{Related Work}\label{sec_related_work}
\subsection{AU-assisted Micro-expression Recognition}
MER has achieved notable progress through advancements in spatiotemporal motion learning. Various methods employ either hand-crafted descriptors or deep learning-based architectures to capture subtle facial dynamics. Some approaches employ algorithmic designs aimed at enhancing local spatiotemporal perception based on the entire face without leveraging AU annotations or AU-based anatomical priors. These methods typically utilize frameworks such as contrastive learning \cite{wang2023temporal, wei2023cmnet}, self-supervised pretext tasks \cite{fan2023selfme, nguyen2023micron, liu2024tgmae}, prototypical embedding strategies \cite{Zhao2022MEPLANAD}, and cascaded refinement pipelines \cite{Zhao2021AT3}. While demonstrating competitive performance, such approaches often lack anatomical grounding, resulting in limited interpretability. 

As a significant human facial expression descriptor, the FACS provides a comprehensive encoding of facial muscle movements through AU annotations, offering valuable guidance for capturing the subtle motion features of MEs. Consequently, many MER methods incorporate AU information during spatiotemporal feature learning, by integrating prior AU knowledge into model design or by directly learning AU motion features to aid ME recognition. Below, we summarize key methodologies based on their use of AU-assisted techniques.

\subsubsection{Feature Extraction from AU-Localized ROIs}
A range of approaches have utilized common observed AUs to localize facial ROIs and extract either hand-crafted or deep learning-based features to capture subtle facial movements. Early works predominantly focused on hand-crafted features followed by a traditional supervised classifier such as SVM for target emotion recognition. For example, Polikovsky et al. \cite{polikovsky2009facial} proposed the 3D Histograms of Oriented gradients (3D-HOG) descriptor to extract the appearance-based ME features from predefined 12 non-overlapping ROIs for MER. Subsequently, a series of hand-crafted features like LBP-TOP \cite{pfister2011recognising}, MDMO \cite{liu2015main} and Sparse MDMO \cite{liu2018sparse} were designed to represent local facial dynamics. These methods laid the foundation for localized feature extraction, but the specially designed features are shallow and low-level, making it difficult to model the spatiotemporal motion well.

With the advancement of deep learning, various techniques have been introduced into MER for significant ME movement representation learning. To better capture the rapid and subtle changes of MEs, several methods leverage the optical flow-based features from AU-localized ROIs for subsequent feature learning \cite{su2021key,wang2024jgulf,Zhao2021AT3}. For instance, Su et al. \cite{su2021key} proposed to utilize face semantic segmentation probability maps as guidance for localizing key facial AU components on extracted optical flow vectors. Similarly, Wang et al. \cite{wang2024jgulf} divided the facial optical flow group features into nine localized regions with multiple scales and overlaps according to common AU appearance. Consequently, they performed joint local and global feature learning by employing the Convolution Neural Network (CNN) and Vision Transformer (ViT) blocks to enhance the MER performance. Still, these methodologies present certain disadvantages. The extracted optical flow-based features are highly susceptible to environmental changes such as lighting, and blinking or disruptive head movements can also impair their performance.

It is worth noting that, directly extracting features from ROIs on RGB images using deep learning methods is another prominent approach. Aouayeb et al. \cite{aouayeb2021micro} used a CNN+LSTM architecture to extract spatiotemporal features from six predefined AU ROIs. Similarly, Deep3DCANN \cite{thuseethan2023deep3dcann} employed 3D CNNs on image sequences, followed by a deep artificial neural network to trace the useful visual associations between different facial sub-regions. Additionally, Li et al. \cite{li2020joint} applies convolutional layers to six ROI regions while considering their contributions to MER. In FRL-DGT \cite{zhai2023feature}, nine AUs are selected and each AU corresponds to multiple ROIs based on the relationship between MEs and facial muscle movements. Then three Transformer-based fusion modules are proposed to extract multi-level features following adaptive displacement generation. However, these methods often ignore global facial information, which is critical for accurately interpreting MEs that involve subtle interactions across the face.

\subsubsection{AU-Relation Modeling via Graph Neural Networks}
Another line of research has explored the utilization of AUs in graph-based representations. These methods typically model the geometric and dynamic interactions between AU regions through GNNs. Lo et al. \cite{lo2020mergcnmr} introduced an AU-assisted Graph Convolution Network (GCN) which first extracted facial AU features with 3D CNNs and then used GCNs to model the dependencies between these regions. Similarly, in \cite{Lei2021MicroexpressionRB}, a facial graph was constructed based on AU co-occurrence frequencies, and a GCN was used to learn the facial dynamics from AU nodes. Other approaches, like \cite{kumar2021microexpressioncb}, \cite{zhang2023adaptive} and \cite{kumar2024uncovering}, all employed facial landmarks to build graph structures, using attention-based graph networks to model both spatial and temporal dynamics of MEs. Overall, while AU relational modeling through GNNs offers a promising way to capture complex interactions between facial regions, their performance is still highly dependent on the quality of the input features and the structure of the graph, making them less robust in real-world scenarios where facial movements are more varied and unpredictable.
\subsection{Contrastive Language-Image Pretraining}
Contrastive Language-Image Pretraining (CLIP) \cite{radford2021learning} has emerged as a powerful framework for multimodal learning, linking visual and textual representations in a shared embedding space by leveraging large-scale paired image-text data. The core strength of CLIP lies in its ability to align visual features with natural language descriptions through a contrastive learning objective, allowing it to generalize across diverse visual tasks without task-specific training \cite{ju2022prompting,nag2022zero,rasheed2023fine}. Recently, efforts have been made to extend CLIP to Dynamic Facial Expression Recognition, resulting in the proposal of DFER-CLIP \cite{zhao2023prompting}. Specifically, DFER-CLIP adapts CLIP’s dual-modality structure by incorporating additional transformer encoders into CLIP image encoder for temporal modeling. Besides, they utilize generalized, pre-generated descriptions for distinct emotions rather than simple class names. Compared to our work, which focuses on the subtle and transient ME movements, we require a more powerful spatiotemporal modeling capability in the vision encoder. Additionally, since MEs are complex and fragmented, general emotion descriptions are likely to misalign with the specific facial ME movements, leading to interference. Therefore, we generate textural descriptions using AU labels that correspond directly to each ME sample.

\begin{table*}[t]
    \caption{A visual comparison of common mixup and cut methods with our LocalStaticFaceMix.}\label{tab_aug}
    \centering
    \setlength{\tabcolsep}{2.5pt}
    \begin{threeparttable}
        \begin{tabular}{ccccccccc} 
        \toprule
        Augmentation & N/A & N/A & N/A & MixUp \cite{zhang2017mixup} & CutOut \cite{devries2017improved} & CutMix \cite{yun2019cutmix} & AugMix \cite{hendrycks2019augmix} & LocalStaticFaceMix\\
        \cmidrule(r){1-9}
        Frame & $v_i^{t}$ & $v_j^{0}$ & $v_j^{t}$ & $v_i^{t}$, $v_j^{t}$ & $v_i^{t}$ & $v_i^{t}$, $v_j^{t}$ & $v_i^{t}$ & $v_i^{t}$, $v_j^{0}$\\
        Image & \begin{minipage}[b]{0.2\columnwidth}
		\raisebox{-.5\height}{\includegraphics[width=\linewidth]{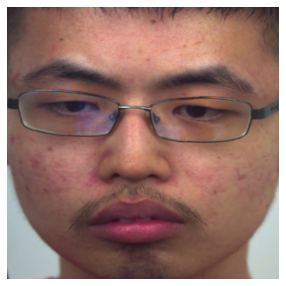}}
	\end{minipage} & \begin{minipage}[b]{0.2\columnwidth}
		\raisebox{-.5\height}{\includegraphics[width=\linewidth]{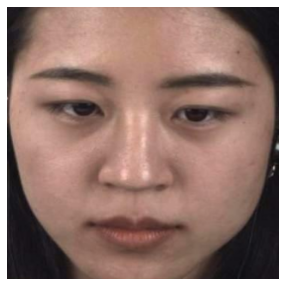}}
	\end{minipage} & \begin{minipage}[b]{0.2\columnwidth}
		\raisebox{-.5\height}{\includegraphics[width=\linewidth]{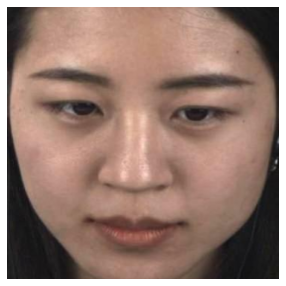}}
	\end{minipage} & \begin{minipage}[b]{0.2\columnwidth}
		\raisebox{-.5\height}{\includegraphics[width=\linewidth]{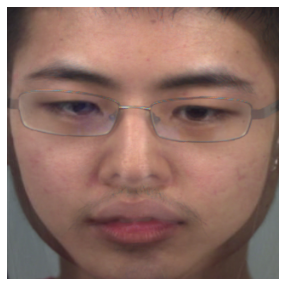}}
	\end{minipage} & \begin{minipage}[b]{0.2\columnwidth}
		\raisebox{-.5\height}{\includegraphics[width=\linewidth]{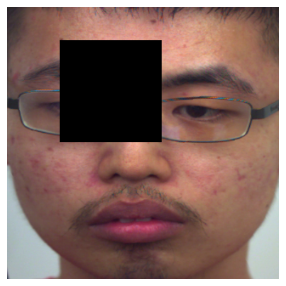}}
	\end{minipage} & \begin{minipage}[b]{0.2\columnwidth}
		\raisebox{-.5\height}{\includegraphics[width=\linewidth]{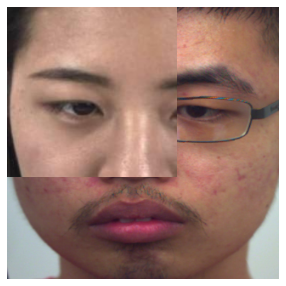}}
	\end{minipage} & \begin{minipage}[b]{0.2\columnwidth}
		\raisebox{-.5\height}{\includegraphics[width=\linewidth]{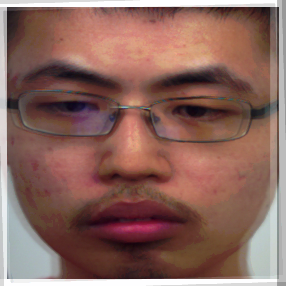}}
	\end{minipage} & \begin{minipage}[b]{0.2\columnwidth}
		\raisebox{-.5\height}{\includegraphics[width=\linewidth]{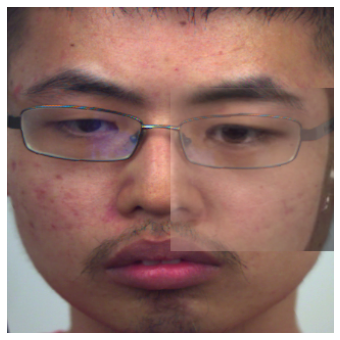}}
	\end{minipage} \\
        \multirow{2}{*}{Label} & \multirow{2}{*}{anger 1.0} & \multirow{2}{*}{neutral} & \multirow{2}{*}{happiness 1.0} & anger 0.5 & \multirow{2}{*}{anger 1.0} & anger 0.6 & \multirow{2}{*}{anger 1.0} & \multirow{2}{*}{anger 1.0}\\
        &&&& happiness 0.5&& happiness 0.4&& \\

        \bottomrule
        \end{tabular}
        \begin{tablenotes}
        \footnotesize
        \item[1] N/A refers to the original images without any data augmentation applied.
        \item[2] $V_i=\{v_i^0, ..., v_i^{T-1}\}$ and $V_j=\{v_j^0, ..., v_j^{T-1}\}$ represent two ME video sequences labeled with \textit{anger} and \textit{happiness}, respectively. $v_j^0$ refers to the onset frame of $V_j$, at which the facial expression begins to occur, approaching a neutral facial state. $v_i^t$ and $v_j^t$ denotes the $t$-th frame in $V_i$ and $V_j$.
        \item[3] The image in the third row is a visualization generated by applying the corresponding augmentation operations to a single frame or a pair of frames from the second row. The labels for the newly generated samples are updated accordingly and noted in the last row.
        \end{tablenotes}
    \end{threeparttable}
    \vspace{-0.4cm}
\end{table*}
\subsection{Facial Micro-expression Data Augmentation}\label{sec_related_work_aug}
Currently, most MER models still face significant data limitations due to the time-consuming and laborious process of ME eliciting and annotation. Common data augmentation techniques like rotating, flipping, and colorjitter are often used to expand MER datasets. However, these traditional methods do little to enhance the diversity of expression variations essential for robust MER, as they cannot simulate unseen faces or change identity information. Advanced augmentation methods like MixUp \cite{zhang2017mixup}, CutMix \cite{yun2019cutmix}, and CutOut \cite{devries2017improved} introduce more diversity by mixing or cutting pixel-level or patch-level information across samples, effectively generating augmented data that enhance generalization ability of models. 
However, as shown in Table \ref{tab_aug}, when applied to the subtle and fragmented nature of MEs, these methods reveal certain limitations. 
Specifically, given two ME video sequences of the same length $V_i=\{v_i^0, ..., v_i^{T-1}\}$ and $V_j=\{v_j^0, ..., v_j^{T-1}\}$, MixUp fuses corresponding frames $v_i^t$ and $v_j^t$ at the pixel level, which obscures the local details and fine-grained movements that are crucial for MEs. CutMix, on the other hand, randomly replaces parts of $v_i^t$ with corresponding regions from $v_j^t$, risking the overwriting of critical ME areas and thereby distorting their subtle characteristics. Additionally, both mixing techniques yield prediction labels as soft labels proportional to the blended regions, yet the complexity of emotional expressions cannot be effectively represented by simple image fusion or patch replacement, let alone by precise proportional mixing of emotions. Finally, the random occlusion applied by CutOut on each frame can mask essential ME regions, such as the mouth corners or eye areas, thus preventing the model from accurately capturing the required facial movements.

In this context, AugMix \cite{hendrycks2019augmix} offers a compelling solution by preserving critical local details while enriching appearance variation. Unlike other methods, AugMix creates diverse images for each frame by applying a series of randomized operations, such as equalize, posterize, rotation, solarize et al, and mixes them in a stochastic manner. Although these transformations introduce subtle distortions in datasets, they are limited to using only the information within a single image, without bringing variations in facial identity.
Therefore, a data augmentation strategy that can enrich facial identity diversity while preserving the integrity of ME movements is in demand.

\section{Method}\label{sec_method}
In this section, we give a detailed explanation of our proposed MER-CLIP model, which is illustrated in Figure \ref{fig:framework}. Firstly, we introduce our designed LocalStaticFaceMix data augmentation strategy. Then we interpret the process of ME video motion encoding. Subsequently, we focus on the two main modules including the AU-Guided Cross-Modal Alignment Module and Emotion Inference Module. Finally, a progressive training strategy is utilized to balance the two modules to enhance the MER performance.

\subsection{Input and Data Augmentation}\label{sec:aug}
\begin{figure}[t] 
    \centering
    \includegraphics[scale=0.37]{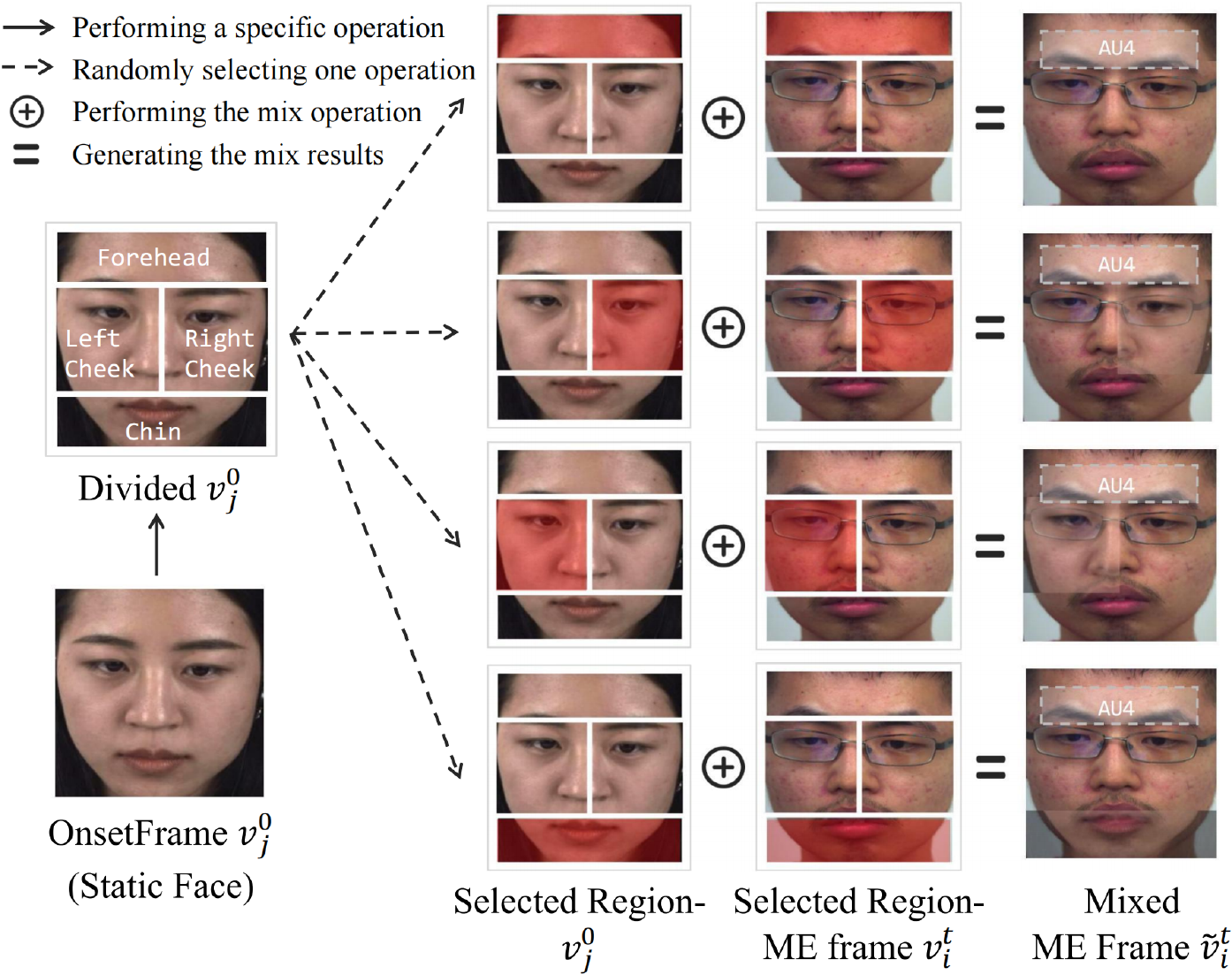}
    \caption{The detailed illustration of proposed LocalStaticFaceMix.}
    \label{fig_mix}
    \vspace{-0.3cm}
\end{figure}

To learn robust spatiotemporal representations, we take a batch of $K$ ME video sequences $\mathcal{V} = \{V_i\}_{i=1}^{K}$ as our model input, where $V_i=\{v_i^0, ..., v_i^{T-1}\}$ and $V_j=\{v_j^0, ..., v_j^{T-1}\}$ represent two different sequences within the batch. During training, these sequences typically undergo a series of data augmentations to enhance generalization and mitigate overfitting, which is particularly crucial given the limited dataset size. However, as analyzed in Section \ref{sec_related_work_aug}, certain effective visual mixing augmentation strategies directly blend expressive facial frames $v_i^t$ and $v_j^t$ when applied to MEs, thereby disrupting the subtle and fragmented motion patterns essential for learning effective motion features. To address this, we propose a novel mixing augmentation strategy tailored for ME sequences, called LocalStaticFaceMix, as illustrated in Figure \ref{fig_mix}.

Specifically, we introduce a strategy that mixes each frame $v_i^t$ from sequence $V_i$ with the onset frame $v_j^0$ of another sequence $V_j$, which can be seen as the neutral static face before any expression occurs \cite{Yan2014CASMEIA}. This design enhances facial appearance diversity by blending two facial images while preserving ME-specific dynamics by avoiding interference between active expression phases.
Furthermore, to prevent global pixel-level blending from obscuring critical local ME features, we partition each input facial image into four distinct regions, defined as the Forehead ($R_f=I[0:\frac{H}{4}, :]$), Chin ($R_c=I[\frac{3H}{4}:H, :]$), Left cheek ($R_{lc}=I[\frac{H}{4}:\frac{3H}{4}, 0:\frac{W}{2}]$), and Right cheek ($R_{rc}=I[\frac{H}{4}:\frac{3H}{4}, \frac{W}{2}:W]$), where $I$ represents any input facial image; $H$ and $W$ are the height and width of the image, respectively.

For each sequence, one of these regions $R\in\{R_f,R_c,R_{lc},R_{rc}\}$ is randomly selected to guide the localized blending operation:
\begin{equation} \begin{split}
    &\tilde{v}_i^t = v_i^t \odot (1 - M_R) + \\
    &[\omega \cdot (v_i^t \odot M_R) + (1 - \omega) \cdot (v_j^0 \odot M_R)],
\end{split} \end{equation} 
where $\omega \in [0,1]$ is the coefficient used to control the ratio of the two mixtures. $M_R$ is a binary mask corresponding to the selected region $R$, and $\odot$ denotes element-wise multiplication.
We also show the detailed operation process of LocalStaticFaceMix in Algorithm \ref{alg:aug}. 

By applying the blending to each frame of a sequence with a partial region of the onset frame from another sequence, LocalStaticFaceMix effectively minimizes disruption of the original ME motion patterns while introducing facial identity variability that reduces overfitting in small datasets. Moreover, through extensive experiments, we found that combining LocalStaticFaceMix with AugMix \cite{hendrycks2019augmix} significantly improves the performance of MER (Section \ref{sec_abla}).

\begin{algorithm}[t]
\label{alg:aug}
\caption{LocalStaticFaceMix for ME Data Augmentation}
\KwIn{A batch of $K$ ME video sequences $\mathcal{V} = \{V_i\}_{i=1}^{K}$, where each sequence $V_i = \{v_i^t\}_{t=0}^{T-1}$ consists of $T$ frames.}
\KwOut{Augmented batch of ME video sequences $\tilde{\mathcal{V}}$.}

\For{each video sequence $V_i = \{v_i^t\}_{t=0}^{T-1}$ in the batch}{
    1. Randomly select the first frame $v_j^0$ from another sequence $V_j = \{v_j^t\}_{t=0}^{T-1}$ within the same batch\;
    2. Partition each frame into four regions: $R_f$ (forehead), $R_c$ (chin), $R_{lc}$ (left cheek), $R_{rc}$ (right cheek)\;
    3. Randomly select one region $R\in\{R_f,R_c,R_{lc},R_{rc}\}$ for mixing and generate a binary mask $M_R$ corresponding to this region\;
    4. \For{each frame $v_i^t$ in $V_i$}{
        Perform the following mixing operation with $v_j^0$:
        \begin{equation} \begin{split}
        &\tilde{v}_i^t = v_i^t \odot (1 - M_R) + \\
        &[\omega \cdot (v_i^t \odot M_R) + (1 - \omega) \cdot (v_j^0 \odot M_R)],
        \end{split} \end{equation} 
        where $\omega \in [0, 1]$ is the mixing ratio and $\odot$ denotes element-wise multiplication\;
    }
    5. Replace the original sequence $V_i$ with the augmented sequence $\tilde{V}_i = \{\tilde{v}_i^t\}_{t=0}^{T-1}$\;
}

\Return Augmented batch $\tilde{\mathcal{V}}$\;
\end{algorithm}


\subsection{Video Motion Encoding}\label{sec:video_encoder}
\begin{figure}[tb] 
    \centering
    \includegraphics[scale=0.37]{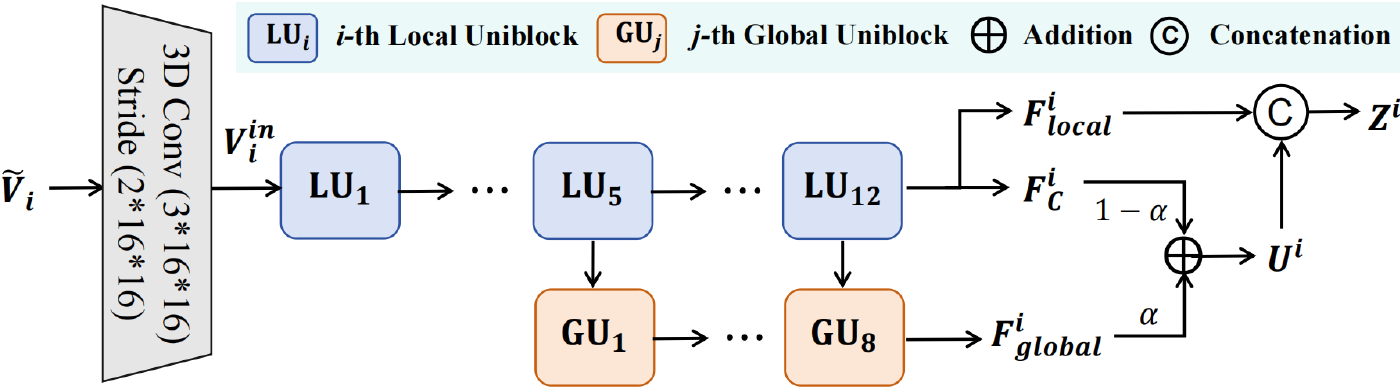}
    \caption{The detailed flowchart of the Video Motion Encoding Process.}
    \label{fig_uf2}
    \vspace{-0.3cm}
\end{figure}

To extract motion features from an augmented input ME video sequence $\tilde{V}_i\in \mathbb{R}^{H\times W\times T}$ (with standardized spatial dimensions $H\times W$, and temporal length $T$), we employ UniformerV2 \cite{li2022uniformerv2} as our vision encoder $f_{vis}$. UniformerV2 is an advanced video encoder built upon image-pretrained Vision Transformers (ViTs). In our implementation, as shown in Figure \ref{fig_uf2}, $\tilde{V}_i$ is first downsampled spatially by a factor of 16 and temporally by a factor of 2 via a 3D convolution layer, resulting in the projection of $\tilde{V}_i$ into $L$ spatiotemporal tokens $V_i^{in} \in \mathbb{R}^{L\times D_1}$, where $L=\frac{H}{16}\cdot\frac{W}{16}\cdot\frac{T}{2}$ and $D_1$ denotes the channel dimension. Subsequently, local and global UniBlocks \cite{li2022uniformerv2} are incorporated to model multi-scale spatiotemporal representations, which are efficient video design modules derived from image-pretrained ViT blocks. Furthermore, the multi-stage fusion of these representations enables the encoder to effectively capture complex temporal dynamics in videos. The above enhancement is particularly beneficial for modeling subtle facial movements essential for MER. 

From the final local UniBlock, we extract both the class token $F_C^i\in\mathbb{R}^{1\times D_1}$ and local visual feature $F_{local}^i \in \mathbb{R}^{L\times D_1}$. These are then dynamically integrated with the final global video representation $F_{global}^i \in \mathbb{R}^{1\times D_1}$ extracted from global UniBlocks. Specifically, we first obtain the globally spatiotemporal motion features by computing the element-wise weighted sum of $F_C^i$ and $F_{global}^i$ as UniformerV2’s design:
\begin{equation}
\begin{aligned}
U^i = \alpha F_{global}^i + (1-\alpha) F_C^i \in \mathbb{R}^{1\times D_1},
\end{aligned}
\end{equation}
where $\alpha$ is a learnable parameter processed by a sigmoid function. The resulting feature $U^i$ is then fed into the AU-Guided Cross-Modal Alignment Module, which enhances video encoder’s ability to capture fine-grained ME movements through the alignment of visual and textual features (see Section \ref{sec:au-guided}). Furthermore, to preserve sufficient spatiotemporal motion characteristics for subsequent emotion inference, we concatenate $U^i$ with $F_{local}^i$ to form a comprehensive feature representation:
\begin{equation}
\begin{aligned}
\label{eq_zi}
Z^i = {\rm Concat}(U^i, F_{local}^i) \in\mathbb{R}^{(1+L)\times D_1}.
\end{aligned}
\end{equation}
The concatenated feature $Z^i$ is then passed to the Emotion Transformer Head, where high-level emotional semantic features are extracted for MER (see Section \ref{sec:emotion}).

\begin{table*}[t]
    \caption{Descriptions of key AUs appearing in micro-expressions.}\label{tab_AU}
    \centering
        \begin{tabular}{ccc|ccc} 
        \toprule
        AU Number & FACS Definition & Action-Oriented Description & AU Number & FACS Definition & Action-Oriented Description\\
        \cmidrule(r){1-3} \cmidrule(r){4-6} 
        AU1 & Inner Brow Raiser & raising the inner part of the brows & AU14 & Dimpler & creating the dimples \\
        AU2 & Outer Brow Raiser & raising the outer part of the brows & AU15 & Lip Corner Depressor & depressing the corners of the lips \\
        AU4 & Brow Lowerer & lowering the brows & AU16 & Lower Lip Depressor & depressing the lower lip \\
        AU5 & Upper Lid Raiser & raising the upper lids & AU17 & Chin Raiser & raising the chin \\
        AU6 & Cheek Raiser & raising the cheeks & AU20 & Lip Stretcher & stretching the lips \\
        AU7 & Lid Tightener & tightening the lids & AU23 & Lip Tightener & tightening the lips \\
        AU9 & Nose Wrinkler & wrinkling the nose & AU24 & Lip Presser & pressing the lips \\
        AU10 & Upper Lip Raiser & raising the upper lip & AU25 & Lips Part & parting the lips \\
        AU12 & Lip Corner Puller & pulling the corners of the lips & AU28 & Lip Suck & sucking the lips \\
        \bottomrule
        \end{tabular}
\end{table*}

\subsection{AU-Guided Cross-Modal Alignment Module}\label{sec:au-guided}
As a general-purpose visual feature extractor, UniformerV2 still faces limitations when capturing fine-grained facial motion cues. To address this, we introduce an AU-guided cross-modal enhancement, which leverages detailed textual descriptions to refine motion encoding.
\subsubsection{Text Encoding with AU-Guided prompts}
For the textual part, we encode a set of prompts generated from AU descriptions and multiple templates using the CLIP-pretrained text encoder $f_{text}$. According to FACS, each AU has an anatomically precise definition. However, the standard FACS terminology lacks intuitive motion semantics, which may hinder effective cross-modal alignment with visual features. To mitigate this issue, we convert each AU into an action-oriented natural language description, which explicitly specifies the location and movement of facial muscles, as illustrated in Table \ref{tab_AU}. 

Given a ME video sequence $V_i$ with associated AU annotations $\mathcal{A}_i=\{AU_m\}_{m=1}^{M_i}$, where $M_i$ is the number of activated AUs in $V_i$ and $AU_m$ represents an activated AU, we construct a comprehensive movement description by concatenating the action-oriented descriptions $d(AU_m)$ of each AU:
\begin{equation}
\begin{aligned}
\mathcal{D}_{i}=\bigotimes_{m=1}^{M_i}(d(AU_m)),
\end{aligned}
\end{equation}
where $\bigotimes$ denotes the ordered concatenation operator. This process yields a detailed and structured ME movement description $\mathcal{D}_{i}$, such as \textit{raising the inner part of the brows} (AU1) or \textit{a combination of raising the cheeks and pulling the corners of the lips} (AU6+AU12).

To further improve the model's comprehension and enrich the semantic context, we manually design multiple textual templates $\mathcal{T}=\{Tem_n\}_{n=1}^N$, each containing a placeholder slot. For example, a template may take the form of
$\textit{“This micro-expression involves \{\}”}$,
or
$\textit{“One can identify this micro-expression by \{\}”}$. During training, we randomly select a template $Tem_n \in \mathcal{T}$ for each sample $V_i$, and its placeholder is populated with the corresponding  $\mathcal{D}_i$, forming a detailed ME textual prompt $\mathcal{P}_i$:
\begin{equation}
\begin{aligned}
\mathcal{P}_i=Tem_n(\mathcal{D}_i).
\end{aligned} \end{equation} 
The completed prompt $\mathcal{P}_i$ is then passed into the CLIP text encoder $f_{text}$, and the output textual feature vector $X_{text}^i$ is obtained via a text projector $g_{text}$:
\begin{equation}
\begin{aligned}
X_{text}^i=g_{text}(f_{text}(\mathcal{P}_i)) \in \mathbb{R}^{1\times D_2}.
\end{aligned} \end{equation} 
The resulting textual feature vector $X_{text}^i$ provides a fine-grained representation of facial muscle movements, which is subsequently aligned with the visual features within a shared embedding space of dimension $D_2$.

\subsubsection{Cross-Modal Alignment}
In AU-Guided Cross-Modal Alignment Module, the globally spatiotemporal visual representation $U^i$ is projected into the embedded space shared with the textual feature vector $X_{text}^i$, yielding $X_{vis}^i$:
\begin{equation} \begin{aligned}
X_{vis}^i=g_{vis}(U^i) \in \mathbb{R}^{1\times D_2},
\end{aligned} \end{equation} 
where $g_{vis}$ is a visual projector. The goal is to align the representations from both modalities so that visual features corresponding to specific ME sequences are paired with their associated AU-guided textual descriptions.
This alignment is achieved using the CLIP contrastive loss $\mathcal{L}_{CLIP}$, which operates on pairs of visual and textual features \cite{radford2021learning}. Specifically, we aim to maximize the cosine similarity between the correct video-text pairs (i.e., matching visual and textual representations) while minimizing the similarity between incorrect pairs. 

Formally, given a batch of $K$ ME video sequences and their corresponding AU prompts, the model obtains the encoded visual features $\{X_{vis}^i\}_{i=1}^{K}$ and textural features $\{X_{text}^i\}_{i=1}^{K}$, both have been projected into a shared embedding space. The cosine similarity between a pair of visual and textual embeddings is given by:
\begin{equation} \begin{aligned}
S(X_{vis}^i,X_{text}^j)=\frac{X_{vis}^i \cdot X_{text}^j}{\|X_{vis}^i\|\|X_{text}^j\|}, 
\end{aligned} \end{equation} 
where $\cdot$ denotes the dot product. The contrastive loss $\mathcal{L}_{CLIP}$ is then defined as:
\begin{equation}
\begin{aligned}
\mathcal{L}_{CLIP}=-\frac{1}{2K}\sum_{i=1}^K(\log\frac{\exp(S(X_{vis}^i,X_{text}^i)/\tau)}{\sum_{j=1}^K \exp(S(X_{vis}^i,X_{text}^j)/\tau)} \\
+\log\frac{\exp(S(X_{vis}^i,X_{text}^i)/\tau)}{\sum_{j=1}^K \exp(S(X_{vis}^j,X_{text}^i)/\tau)}),
\end{aligned}
\end{equation}
where $\tau$ is a temperature parameter. This loss function encourages high similarity for the matching pairs $(X_{vis}^i,X_{text}^i)$, while penalizing the similarities for mismatched pairs $(X_{vis}^i,X_{text}^j)$ where $i\neq j$.
By optimizing this contrastive loss, the model learns to associate specific visual cues from MEs with their corresponding textual motion descriptions based on AU labels, thereby improving the vision encoder to capture subtle facial ME movements and learn spatiotemporal representations.
\subsection{Emotion Inference Module}\label{sec:emotion}

In this module, we address the semantic gap between motion features and emotion classification, refining the learned representations for final ME recognition.
Specifically, instead of the conventional linear classification head, we introduce a lightweight transformer head $h_{emo}$ after the vision encoder $f_{vis}$. This allows the model to further mine emotional meanings from ME motion patterns and extract emotional features with a higher level of semantic understanding.

In terms of architectural composition, $h_{emo}$ consists of two transformer blocks followed by a linear layer. For each input ME sequence $V_i$, the comprehensive ME movement representation $Z^i$ obtained by Eq. \ref{eq_zi}, is fed into $h_{emo}$, obtaining the predicted emotion scores $\mathbf{s}^i$:
\begin{equation} \begin{aligned}
\mathbf{s}^i=h_{emo}(Z^i) \in \mathbb{R}^C,
\end{aligned} \end{equation} 
where $C$ is the number of emotion classes. These scores are converted into probabilities via a softmax operation, resulting in the predicted probability distribution $\mathbf{p}^i$:
\begin{equation} \begin{aligned}
\mathbf{p}^i = {\rm softmax}(\mathbf{s}^i) \in [0,1]^C.
\end{aligned} \end{equation} 

To address the common class imbalance in ME datasets, we apply the focal loss \cite{lin2017focal} as the classification loss $\mathcal{L}_{CLS}$, which is formulated as:
\begin{equation} \begin{aligned}
\mathcal{L}_{CLS}(\mathbf{y}, \mathbf{p})=-\frac{1}{K} \sum_{i=1}^K \sum_{c=1}^C y^{i,c}(1-p^{i,c})^\gamma\log(p^{i,c}).
\end{aligned} \end{equation} 
$\mathbf{y}^i \in \{0, 1\}^C$ is the ground-truth one-hot label vector, where $y^{i,c}=1$ if the sample $i$ belongs to class $c$, and $0$ otherwise. $p^{i,c}$ is the predicted probability for class $c$. $\gamma$ is the focusing parameter emphasizing hard-to-classify examples, which is set to 2 by default. $K$ is the number of samples in the batch.
\subsection{Progressive Training Strategy}\label{sec:progressive}
To balance the impact of the two modules on model training, we adopt a progressive training strategy, ensuring that the model first learns motion features and gradually incorporates emotion semantics. Specifically, we start training with only $\mathcal{L}_{CLIP}$, prioritizing the alignment between motion features and AU-based textual descriptions. At this stage, the model focuses on learning the spatiotemporal ME motion dynamics. As training progresses, we gradually increase the weight of the classification loss $\mathcal{L}_{CLS}$ and reduce the other one until the ratio of the two losses reaches balance, shifting the model's focus toward learning emotional representations. Therefore, the total loss of the model $\mathcal{L}$ can be formalized as follows,
\begin{equation} \begin{aligned}
\mathcal{L} = \lambda(ep) * \mathcal{L}_{CLIP} + (\lambda_s-\lambda(ep))*\mathcal{L}_{CLS}.
\end{aligned} \end{equation} 
Here we fix the sum of the weights of the two losses $\lambda_s$ to $2.0$. $\lambda (ep)$ refers to the weight of $\mathcal{L}_{CLIP}$ in the $ep$-th epoch, which follows a linear decreasing strategy during training, and can be formalized as:
\begin{equation} \begin{aligned}
\lambda(ep) = \lambda_s - (\lambda_s - \lambda_0) * (ep/EP),
\end{aligned} \end{equation} 
where $\lambda_0$ represents the minimum weight of $\mathcal{L}_{CLIP}$, and here we take a value of $1.0$. $EP$ is the total number of training epochs. In this way, the progressive strategy enables the model to first master the subtle motion patterns in MEs before refining its emotional recognition abilities.
\section{Experiment}\label{sec_experiment}

In this section, we sequentially introduce our experimental settings including the used datasets, evaluation protocols and implementation details. Subsequently, we present a comparison of experimental results between our method and existing state-of-the-art methods across multiple classification tasks on various ME datasets. Finally, the comprehensive ablation studies and interpretability analyses are provided.

\begin{table*}[t]
    \caption{Distributions of emotional labels for different classification tasks on different ME datasets.}\label{tab_dataset}
    \centering
    \begin{threeparttable}
        \begin{tabular}{cccc} 
        \toprule
        Dataset & Classification Task & Total & Distribution of Emotion Labels\\
        \cmidrule(r){1-4}
        \multirow{2}{*}{SAMM} & 5-class & 136 & Anger (57) Contempt (12) Happiness (26) Surprise (15) Others (26) \\
        &3-class & 133 & Negative (92) Positive (26) Surprise (15) \\
        \cmidrule(r){1-4}
        \multirow{2}{*}{CASME II} & 5-class & 249 & Disgust (63) Happiness (32) Repression (27) Surprise (28) Others (99) \\
        &3-class & 156 & Negative (96) Positive (32) Surprise (28) \\
        \cmidrule(r){1-4}
        \multirow{3}{*}{CAS(ME$)^3$} & 7-class & 860 & Anger (64) Disgust (250) Fear (86) Happiness (55) Sadness (57) Surprise (187) Others (161) \\
        &4-class & 860 & Negative (457) Positive (55) Surprise (187) Others (161) \\
        &3-class & 699 & Negative (457) Positive (55) Surprise (187) \\
        \cmidrule(r){1-4}
        \multirow{3}{*}{DFME} & 7-class (train) & 1,856 & Anger (161) Contempt (100) Disgust (548) Fear (265) Happiness (206) Sadness (278) Surprise (298)\\
        & 7-class (test A) & 474 & Anger (39) Contempt (34) Disgust (129) Fear (62) Happiness (63) Sadness (46) Surprise (101)\\
        &7-class (test B) & 299 & Anger (41) Contempt (37) Disgust (58) Fear (38) Happiness (42) Sadness (35) Surprise (48)\\
        \bottomrule
        \end{tabular}
        \begin{tablenotes}
        \footnotesize
        \item[1] Negative is a merge of Anger, Contempt, Disgust, Fear, Repression, and Sadness. Positive is made up of Happiness.
        \item[2] The statistical data of DFME is based on the publicly available part.
        \end{tablenotes}
    \end{threeparttable}
    \vspace{-0.5cm}
\end{table*}

\subsection{Datasets}
To ensure fair and comprehensive evaluation, we conduct extensive experiments on four widely-used ME datasets, each providing detailed emotion label, Action Unit (AU), and key frame (\textit{onset}, \textit{apex}, and \textit{offset}) annotations. Table \ref{tab_dataset} summarizes the classification tasks and associated distributions of emotion labels across various ME datasets in our experiments.

\textbf{SAMM} \cite{Davison2018SAMMAS} includes 159 ME samples from 32 participants across 13 different ethnicities. The videos are recorded at 200 fps with a resolution of 2040$\times$1088.  Eight emotion categories are labeled in SAMM, including \textit{happiness}, \textit{contempt}, \textit{disgust}, \textit{surprise}, \textit{fear}, \textit{anger}, \textit{sadness}, and \textit{others}. For our experiment, we follow the convention of most methods, using both a 5-class setup and a 3-class setup.

\textbf{CASME II} \cite{Yan2014CASMEIA} provides a frame rate of 200fps and facial area resolution of 280$\times$340 to capture more subtle changes in MEs. It contains 255 ME samples from 26 participants in the officially released version. Among these, 249 samples belong to five major emotion classes: \textit{happiness}, \textit{disgust}, \textit{surprise}, \textit{repression}, and \textit{others}, which are used for our 5-class MER task. For the 3-class setup, all samples are grouped into \textit{negative}, \textit{positive}, and \textit{surprise}.

\textbf{CAS(ME$)^3$} \cite{li2022cas} is divided into PART A and PART C based on different elicitation paradigms. In our experiments, we primarily utilize PART A, which includes 860 ME samples. The videos are recorded at 30 fps with a resolution of 1280$\times$720 and annotated with seven emotion labels, including \textit{happiness}, \textit{disgust}, \textit{surprise}, \textit{fear}, \textit{anger}, \textit{sadness}, and \textit{others}. Consistent with the original paper, we further conduct experiments using a 4-class setup and a 3-class setup.


\textbf{DFME} \cite{zhao2023dfme} is currently the largest ME dataset, containing 7,526 ME samples from 656 participants. The dataset is collected using a high-speed camera, capturing ME clips at various high frame rates (500 fps, 300 fps, and 200 fps). Each sample is annotated with emotion labels, including the seven basic emotions (\textit{happiness}, \textit{disgust}, \textit{contempt}, \textit{surprise}, \textit{fear}, \textit{anger}, \textit{sadness}) as well as the category \textit{others}. For our experiment, we utilize its publicly available parts, which have been divided into a training set (1,856 samples), a testing A set (474 samples), and a testing B set (299 samples) \cite{dfme2024ccac}. It is worth noting that different parts of the sample have no overlapping facial identity information.
\vspace{-0.5cm}
\subsection{Evaluation Protocols and Metrics}
In line with the original papers of the datasets and previous works, we adopt the Leave-One-Subject-Out (LOSO) cross-validation strategy for evaluation on the SAMM, CASME II, and CAS(ME$)^3$ datasets. Specifically, for each fold, all samples from one subject are used as the testing set, while the remaining samples are used for training. For the DFME dataset, we follow the evaluation protocols in \cite{dfme2024ccac}, conducting experiments based on the divided training and testing sets.

Given the inherent class imbalance present in most ME datasets, especially for the 3-class MER tasks, we evaluate our model using not only the common accuracy (ACC) metric but also Unweighted F1-score (UF1) and Unweighted Average Recall (UAR) to provide a more balanced assessment of performance. Specifically, ${TP}_c$, ${FP}_c$, and ${FN}_c$ are true positives, false positives, and false negatives of class $c$, respectively. Then we can calculate as follows:
\begin{equation} \begin{aligned}
{\rm UF1} = \frac{1}{C} \sum_{c=1}^{C} \frac{2\cdot {\rm Precision}_c \cdot {\rm Recall}_c}{{\rm Precision}_c + {\rm Recall}_c},
\end{aligned} \end{equation} 
where ${\rm Precision}_c$ and ${\rm Recall}_c$ can be formulized as:
\begin{equation} \begin{aligned}
{\rm Precision}_c = \frac{{TP}_c}{{TP}_c+{FP}_c}, 
\end{aligned} \end{equation} 
\begin{equation} \begin{aligned}
{\rm Recall}_c = \frac{{TP}_c}{{TP}_c+{FN}_c}.
\end{aligned} \end{equation} 
UF1 provides the average F1-score across all classes without weighting by class frequency.
\begin{equation} \begin{aligned}
{\rm UAR} = \frac{1}{C} \sum_{c=1}^{C} {\rm Recall}_c.
\end{aligned} \end{equation} 
UAR averages the recall of each class, treating all classes equally, regardless of their frequency.
\vspace{-0.3cm}
\subsection{Implementation Details}
\subsubsection{Model Configuration}
For the vision encoder, we utilize the UniformerV2 \cite{li2022uniformerv2} with the CLIP-ViT architecture as the backbone. To enhance temporal modeling, we insert global UniBlocks into the last 8 layers of ViT-B/16. It is worth noting that we initialize UniformerV2 with open-source parameters pre-trained on SSV2 \cite{goyal2017something}. Concerning the textual part, we employ the pre-trained CLIP text encoder \cite{radford2021learning}, where the maximum number of textual tokens is set to 77. During the training process, the CLIP text encoder is frozen, and we finetune the entire UniformerV2. Additionally, the Emotion Transformer Head is learned from scratch, which consists of two Transformer blocks with a hidden size of 512 and a fully connected layer for classification. For the CLIP loss $\mathcal{L}_{CLIP}$, the temperature hyper-parameter $\tau$ is set to 0.07 to balance contrastive learning effectively.

\begin{table}[t]
\caption{Comparison of experimental results with other methods for 3-class classification on SAMM and CASME II.}
\centering
\label{tab:result_3casa}
\begin{threeparttable}
\begin{tabular}{c c c c c}
  \toprule
  \multirow{2}{*}{MER Methods} & \multicolumn{2}{c}{CASME II} & \multicolumn{2}{c}{SAMM} 
  \\
  \cmidrule(r){2-3}\cmidrule(r){4-5}
  & UF1 & UAR & UF1 & UAR \\
  \cmidrule(r){1-5}
  OFF-ApexNet (2019) \cite{gan2019off} & 0.8697 & 0.8828 & 0.5049 & 0.5392\\

  STSTNet (2019) \cite{liong2019shallowTS} & 0.8382 &0.8686 & 0.6588 & 0.6810 \\
  
  $\mu$-bert (2023) \cite{nguyen2023micron} & 0.9034 & 0.8914 & N/A & N/A \\
  HTNet (2024) \cite{wang2024htnet} & \textbf{0.9532} & \textbf{0.9516} & \underline{0.8131} & \underline{0.8124} \\
  \cmidrule(r){1-5}
  
  MER-CLIP (ours)& \underline{0.9409} & \underline{0.9487} & \textbf{0.8321} & \textbf{0.8434}\\
  
  \bottomrule
\end{tabular}
\end{threeparttable}
\end{table}

\begin{table}[t]
\caption{Comparison of experimental results with other methods for 5-class classification on SAMM and CASME II.}
\centering
\label{tab:result_5casa}
\begin{threeparttable}
\begin{tabular}{c c c c c}
  \toprule
  \multirow{2}{*}{MER Methods} & \multicolumn{2}{c}{CASME II} & \multicolumn{2}{c}{SAMM} 
  \\
  \cmidrule(r){2-3}\cmidrule(r){4-5}
  & ACC & UF1 & ACC & UF1 \\
  \cmidrule(r){1-5}
  LBP-TOP (2011) \cite{pfister2011recognising} & 0.3968 & 0.3589& 0.3556 & 0.3589\\

  MDMO (2015) \cite{liu2015main} & 0.5169 &0.4966 & N/A & N/A \\
  \cmidrule(r){1-5}
  GraphTCN (2020) \cite{lei2020novel} & 0.7398 & 0.7246 & 0.7500 & 0.6985\\

  AUGCN (2021) \cite{Lei2021MicroexpressionRB} & 0.7427 & 0.7047& 0.7426 & 0.7045 \\
  
  MERSiamC3D (2021) \cite{Zhao2021AT3} & 0.8189 & 0.8300 & 0.6875 & 0.6400\\
  
  KFC-MER (2021) \cite{su2021key} & 0.7276 & 0.7375 & 0.6324 & 0.5709\\

  MERASTC (2021) \cite{gupta2021merastc} & \textbf{0.8540} & \textbf{0.8620} & \textbf{0.8380} & \textbf{0.8440} \\

  MER-Supcon (2022) \cite{zhi2022micro} & 0.7358 & 0.7286 & 0.6765 & 0.6251\\

  C3DBed (2023) \cite{pan2023c3dbed} & 0.7764 & 0.7520 & 0.7573 & 0.7216 \\
  
  $\mu$-bert (2023) \cite{nguyen2023micron} & N/A & \underline{0.8553} & N/A & \underline{0.8386} \\
  
  JGULF (2024) \cite{wang2024jgulf} & 0.8204 & 0.8077 & \underline{0.8071} & 0.7203 \\
  \cmidrule(r){1-5}
  
  MER-CLIP (ours)& \underline{0.8233} & 0.8378 & 0.7721 & 0.7414\\
  
  \bottomrule
\end{tabular}
\end{threeparttable}
\end{table}

\subsubsection{Training Details}
The input to our model comprises pairs of trimmed ME video sequences and the corresponding textual AU prompts. To mitigate background noise and enhance focus on the facial regions, each frame is first aligned and cropped using facial landmark detection \cite{king2009dlib,song2019recognizing}. Following this, we downsample each video sequence to 16 frames by leveraging the key-frames sequence extraction method from \cite{Zhao2022MEPLANAD}, which employs a temporally adaptive sampling strategy. This approach ensures that the onset frame, apex frame, and offset frame are retained, thereby effectively preserving the essential ME  motion characteristics during downsampling.

After the above data preprocessing, the video frames are sent to the model and further center-cropped to 224$\times$224 pixels, and undergo a sequence of data augmentation techniques to improve the robustness and generalization of the model. Specifically, the augmentations are randomly performed with a 50\% probability, including color jitter, horizontal flipping, AugMix (magnitude=1), and our LocalStaticFaceMix (with mixing coefficient $\omega=0.5$).

For model training, the initial learning rate is set to 5e-5. A warmup phase is implemented for the first 5 epochs, followed by a CosineAnnealingLR schedule to adaptively reduce the learning rate until the end of 55 training epochs. The model is optimized using the AdamW optimizer with a weight decay parameter set to 0.05. We implement the whole experiment on two
NVIDIA RTX A6000 GPU with PyTorch toolbox.

\begin{table}[t]
\caption{Comparison of experimental results with other methods for 3-class, 4-class, and 7-class classification on CAS(ME$)^3$.}
\centering
\label{tab:result_casme3}
\begin{threeparttable}
\begin{tabular}{c c c c}
  \toprule
  \multirow{2}{*}{MER Methods} & \multirow{2}{*}{Classification Task} & \multicolumn{2}{c}{CAS(ME$)^3$}
  \\
  \cmidrule(r){3-4}
  & & UF1 & UAR\\
  
  \cmidrule(r){1-4}
  
  STSTNet (2019) \cite{liong2019shallowTS} & 3 & 0.3795 &0.3792\\
  RCN-A (2020) \cite{xia2020revealing} & 3 & 0.3928 &0.3893 \\
  FearRef (2021) \cite{zhou2022feature} & 3 & 0.493 &0.3413\\
  $\mu$-bert (2023) \cite{nguyen2023micron} & 3 & 0.5604 & \underline{0.6125} \\
  HTNet (2024) \cite{wang2024htnet} & 3 & \underline{0.5767} & 0.5415\\
  \textbf{MER-CLIP} (ours) & 3 & \textbf{0.7832} & \textbf{0.7606}\\
  
  \cmidrule(r){1-4}

  AlexNet (2022) \cite{zhang2022review} & 4 & 0.2915 &0.2910 \\
  AlexNet + Depth (2022) \cite{zhang2022review} & 4 & 0.3001 &0.2982\\
  $\mu$-bert (2023) \cite{nguyen2023micron} & 4 & \underline{0.4718} & \underline{0.4913} \\
  SFAMNet (2024) \cite{liong2024sfamnet} & 4 & 0.4462 &0.4797\\
  \textbf{MER-CLIP} (ours) & 4 & \textbf{0.6544} & \textbf{0.6242}\\

  \cmidrule(r){1-4}
  
  AlexNet (2022) \cite{zhang2022review} & 7 & 0.1759 & 0.1801 \\
  AlexNet + Depth (2022) \cite{zhang2022review} & 7 & 0.1773 & 0.1829 \\
  $\mu$-bert (2023) \cite{nguyen2023micron}  & 7 & \underline{0.3264} & \underline{0.3254} \\
  SFAMNet (2024) \cite{liong2024sfamnet}  & 7 & 0.2365 & 0.2373 \\
  \textbf{MER-CLIP} (ours) & 7 & \textbf{0.4997} & \textbf{0.5014}\\
  
  \bottomrule
\end{tabular}
\end{threeparttable}
\end{table}

\begin{table}[t]
\caption{Comparison of experimental results with other methods for 7-class classification on the testA and testB of DFME dataset.}
\centering
\label{tab:result_dfme}
\begin{threeparttable}
\begin{tabular}{c c c c c}
  \toprule
  \multirow{2}{*}{MER Methods} & \multirow{2}{*}{Test Set} & \multicolumn{3}{c}{DFME}
  \\
  \cmidrule(r){3-5}
   & & UF1 & UAR & ACC\\
  
  \cmidrule(r){1-5}
  FearRef (2021) \cite{zhou2022feature} & Test A & 0.3410 &0.3686 & \underline{0.5084}\\
  Wang el al (2024) \cite{dfme2024ccac} & Test A & 0.4067 &0.4074 & 0.4641\\
  He el al (2024) \cite{dfme2024ccac} & Test A & \underline{0.4123} & \underline{0.4210} & 0.4873 \\
  \textbf{MER-CLIP} (ours) & Test A & \textbf{0.5024} & \textbf{0.5115} & \textbf{0.5696} \\
  
  \cmidrule(r){1-5}
  FearRef (2021) \cite{zhou2022feature} & Test B & 0.2875 & 0.3228 & 0.3645\\
  Wang el al (2024) \cite{dfme2024ccac} & Test B & 0.3534 & 0.3661 & 0.3813\\
  He el al (2024) \cite{dfme2024ccac} & Test B & \underline{0.4016} & \underline{0.4008} & \underline{0.4147} \\
  \textbf{MER-CLIP} (ours) & Test B & \textbf{0.5128} & \textbf{0.5120} & \textbf{0.5250}\\
  
  \bottomrule
\end{tabular}
\end{threeparttable}
\end{table}

\subsection{Comparison with State-of-the-Art Methods}

In this section, we present a comprehensive comparison of our proposed MER-CLIP with state-of-the-art methods across multiple ME datasets, focusing on different classification tasks. The results showcase the competitive performance of MER-CLIP, especially on the CAS(ME$)^3$ and DFME datasets, which contain a relatively large number of samples, making the superiority of our model more pronounced.

\begin{figure*}[t]
\centering

\subfloat[CASME II (3-class)]{\includegraphics[width=2.0in]{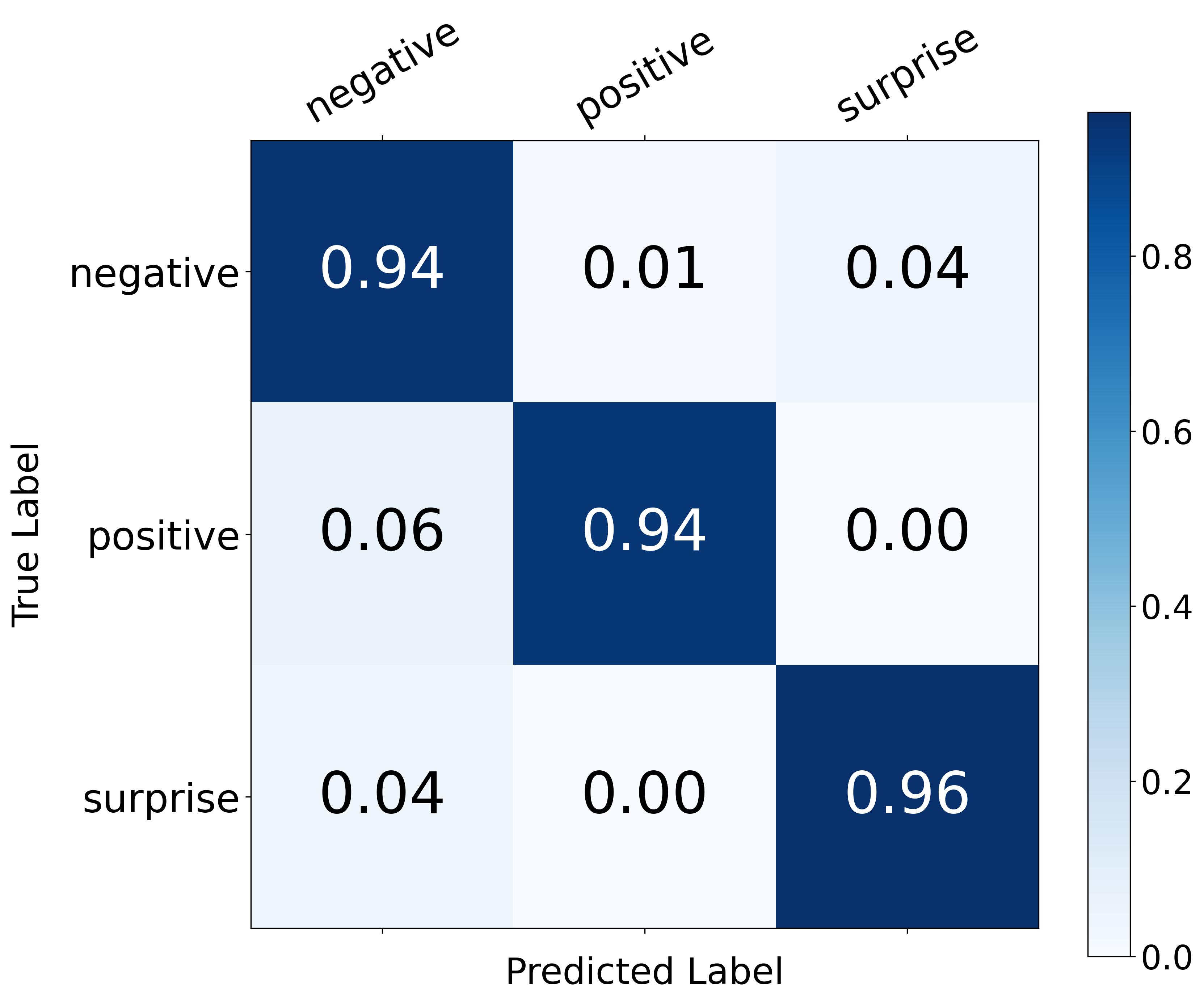}%
\label{fig_casme2_3}}\hspace{3mm}
\subfloat[SAMM (3-class)]{\includegraphics[width=2.0in]{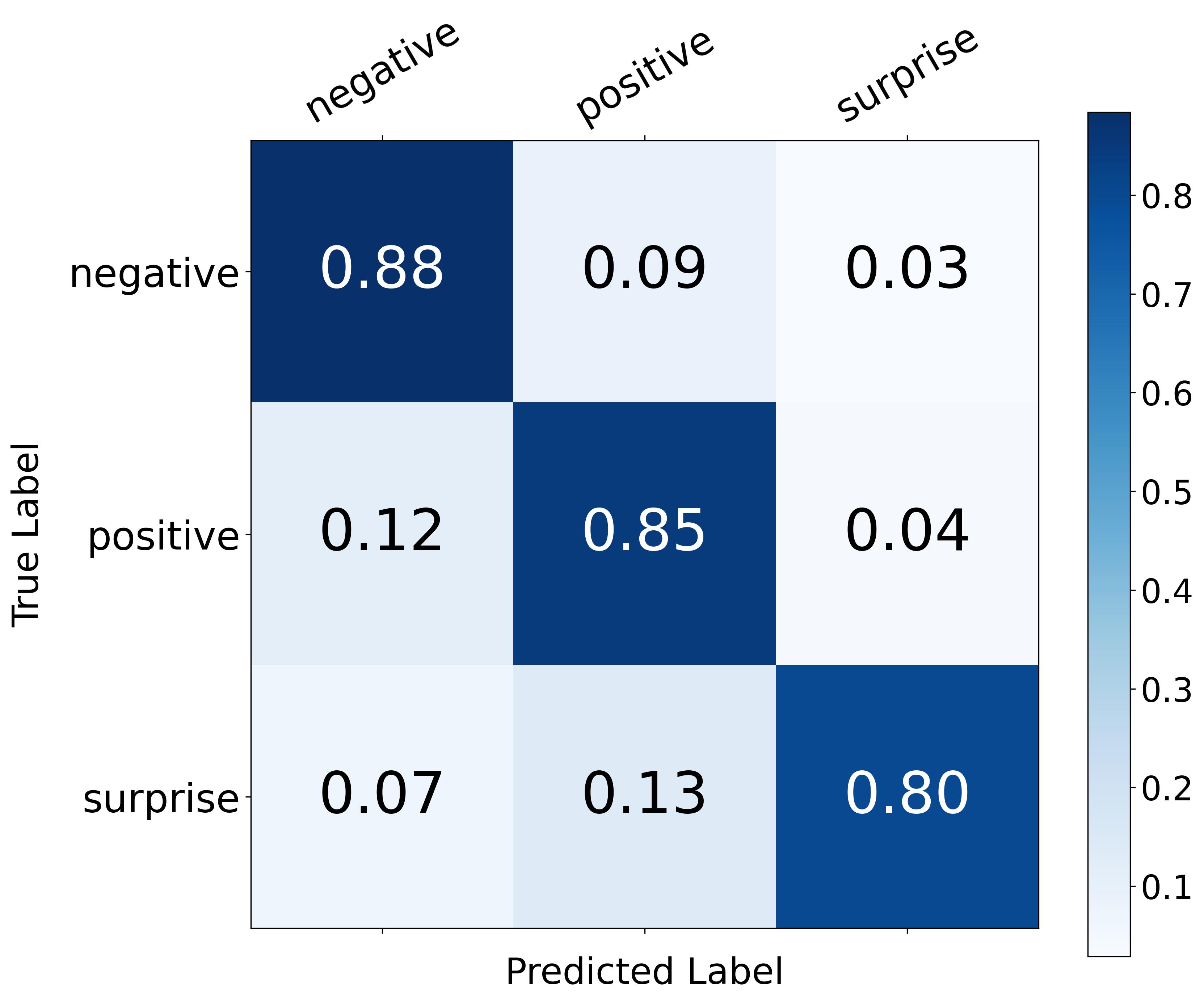}%
\label{fig_samm_3}}\hspace{3mm}
\subfloat[CAS(ME$)^3$ (3-class)]{\includegraphics[width=2.0in]{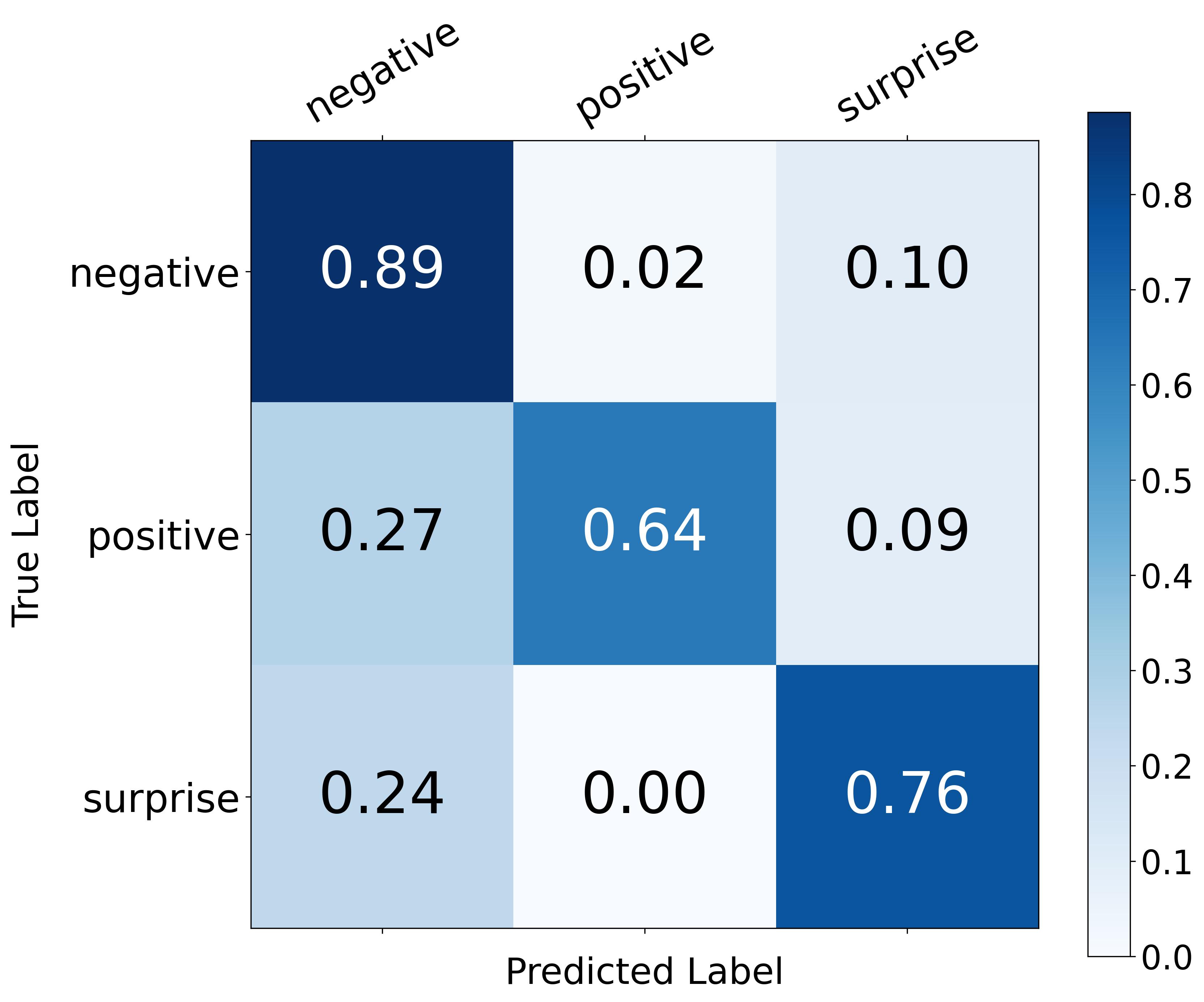}%
\label{fig_casme3_3}}

\subfloat[CASME II (5-class)]{\includegraphics[width=2.0in]{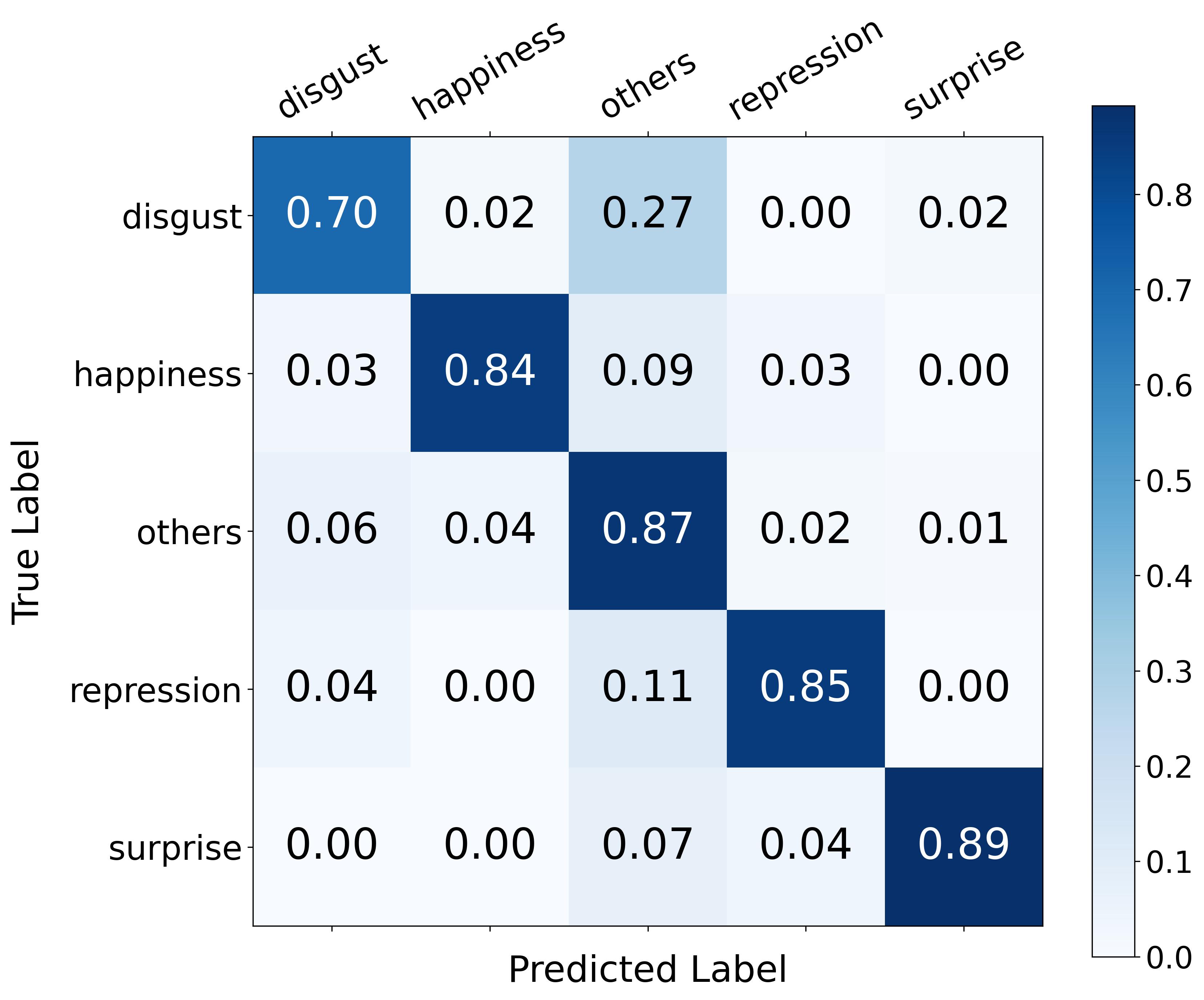}%
\label{fig_casme2_5}}\hspace{3mm}
\subfloat[SAMM (5-class)]{\includegraphics[width=2.0in]{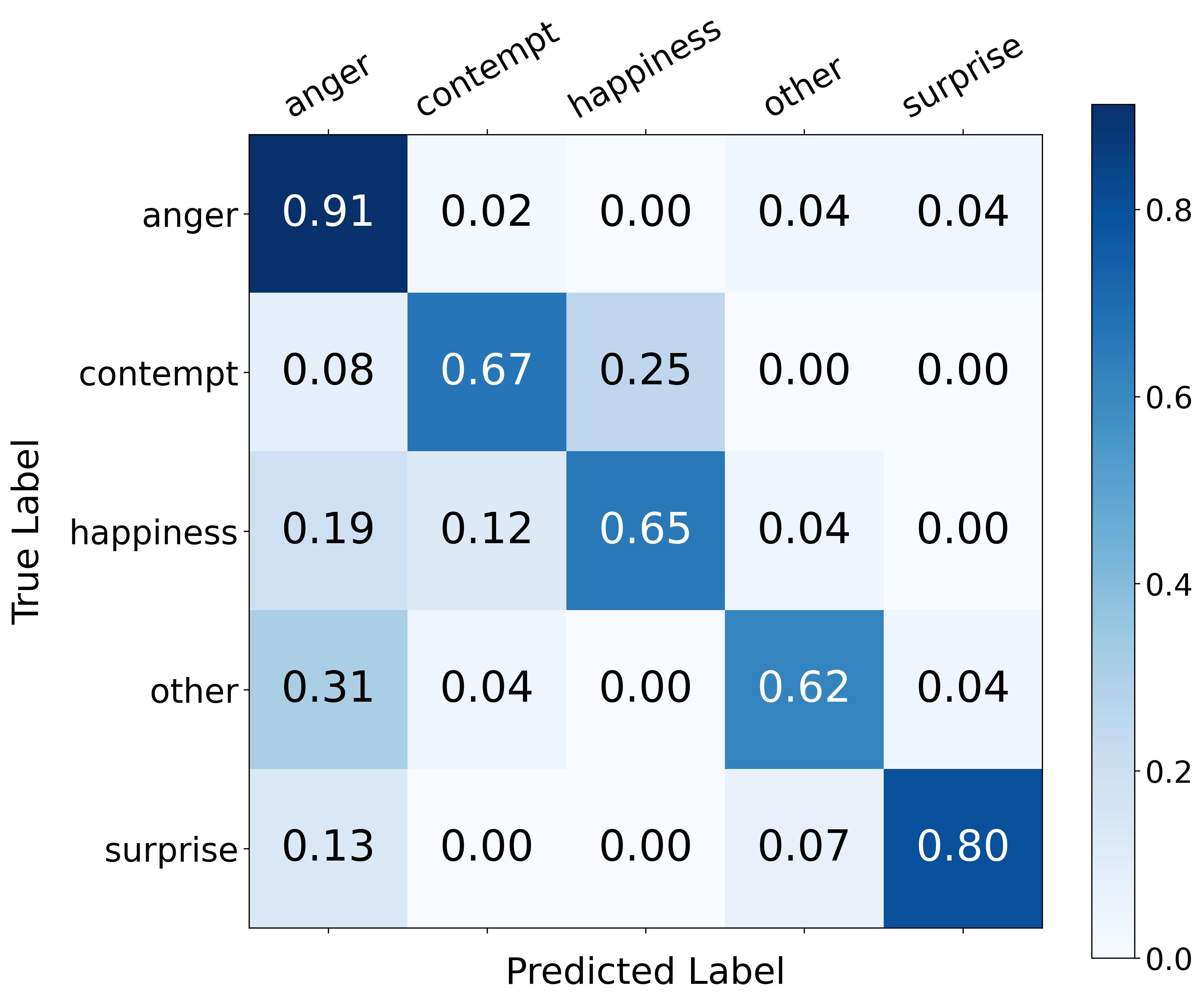}%
\label{fig_samm_5}}\hspace{3mm}
\subfloat[CAS(ME$)^3$ (4-class)]{\includegraphics[width=2.0in]{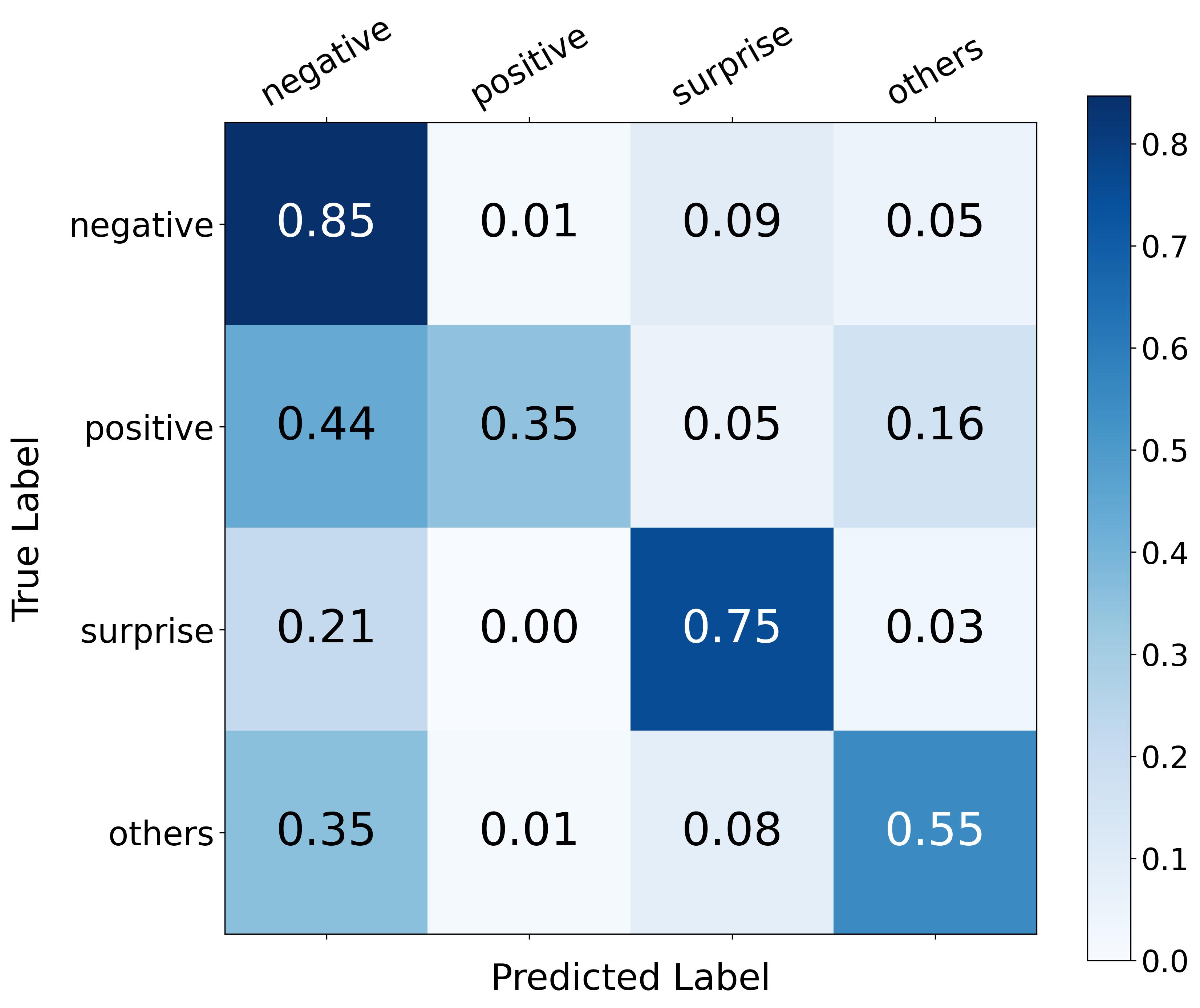}%
\label{fig_casme3_4}}

\subfloat[CAS(ME$)^3$ (7-class)]{\includegraphics[width=2.0in]{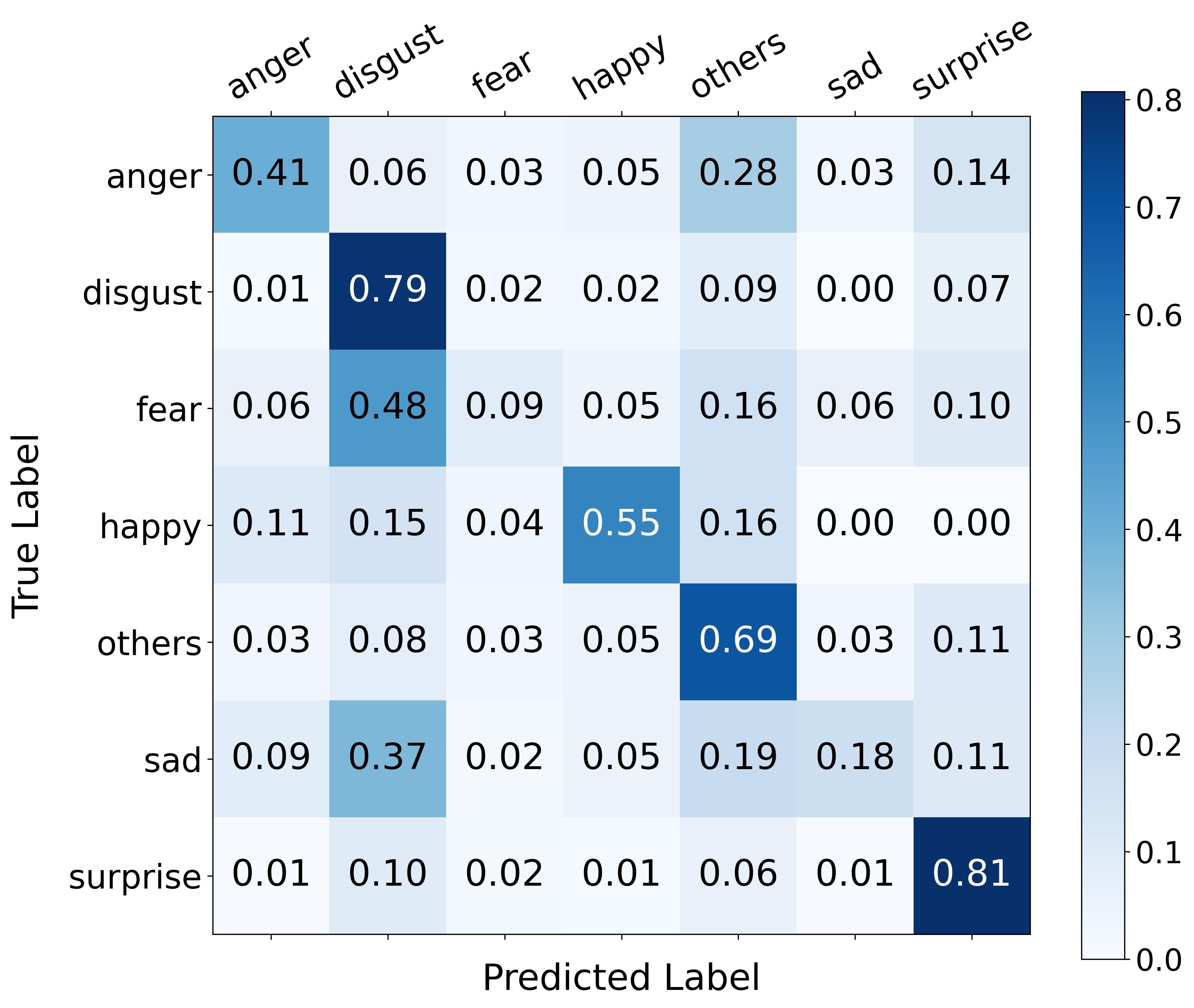}%
\label{fig_casme3_7}}\hspace{3mm}
\subfloat[DFME TestA (7-class)]{\includegraphics[width=2.0in]{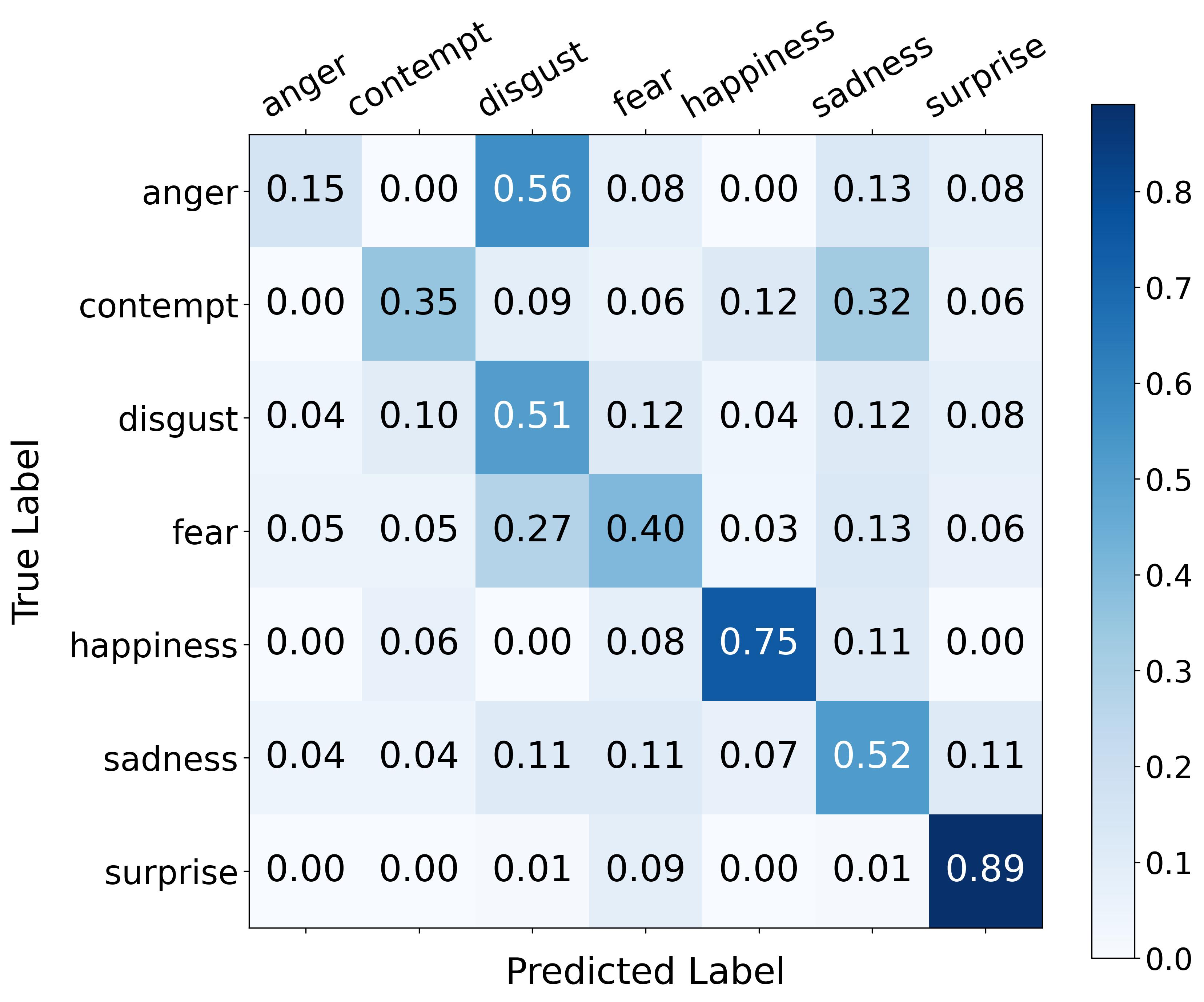}%
\label{fig_dfmea}}\hspace{3mm}
\subfloat[DFME TestB (7-class)]{\includegraphics[width=2.0in]{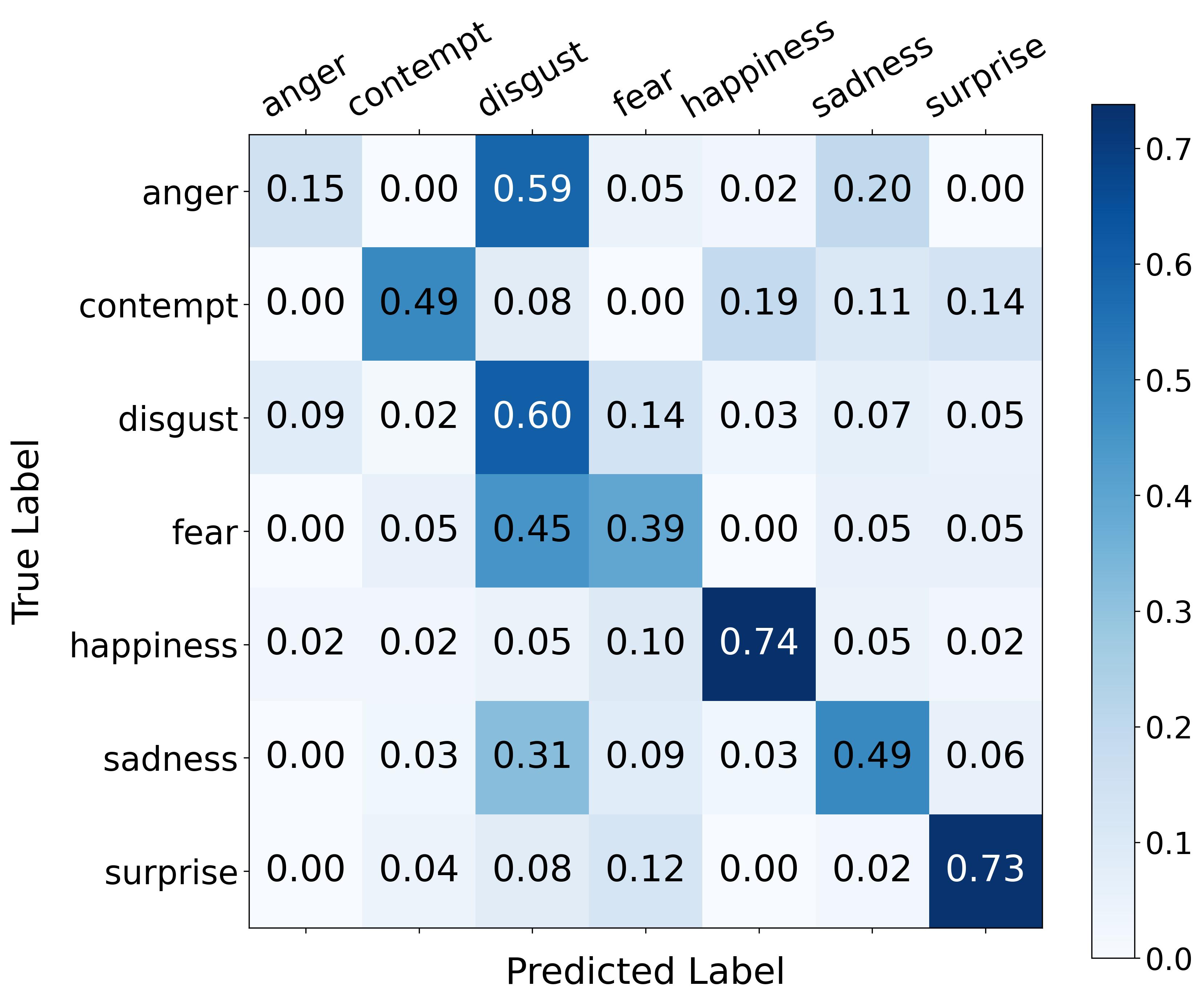}%
\label{fig_dfmeb}}
\caption{Confusion matrices of MER tasks on CASME II, SAMM, CAS(ME$)^3$ and DFME datasets.}
\label{fig_cm}
\end{figure*}

In experiments on the SAMM and CASME II datasets, we conduct both 3-class and 5-class emotion classification tasks to comprehensively assess model performance. Specifically, MER-CLIP demonstrates particularly strong performance in the 3-class task, as illustrated in Table \ref{tab:result_3casa}. On the SAMM dataset, MER-CLIP achieves the highest UF1 score of 0.8424 and UAR of 0.8434, surpassing recent methods such as HTNet \cite{wang2024htnet}. When tested on the CASME II dataset, MER-CLIP obtains a UF1 of 0.9409 and UAR of 0.9487, ranking second only to HTNet. These results highlight the model's effectiveness even on datasets with only about one hundred ME samples.
For the more challenging 5-class task, as shown in Table \ref{tab:result_5casa}, although MER-CLIP does not exceed the top performance of MERASTC \cite{gupta2021merastc}, it remains highly competitive with recent methods like C3DBed \cite{pan2023c3dbed} and JGULF \cite{wang2024jgulf}. MERASTC effectively leverages local facial region appearance features, showing strong results in multi-class tasks on smaller datasets. Due to severe class imbalance and the limited scale of available data, our model—which uses video inputs and a larger parameter set—faces difficulties in achieving distinctly superior performance on the 5-class task. However, the confusion matrices in Figure \ref{fig_cm} provide a more intuitive view of our MER performance. For example, in the 5-class task, the category \textit{others} frequently overlaps with other classes due to its inherently complex label, encompassing composite emotions that are difficult to distinguish. Similarly, in the SAMM dataset’s 5-class task, there is notable confusion between \textit{contempt} and \textit{happiness}, likely because \textit{contempt} often includes elements of ridicule or sarcasm, sharing some features with \textit{happiness}.

For the CAS(ME)$^3$ dataset, we evaluate MER-CLIP on 3-, 4-, and 7-class classification tasks, as shown in Table \ref{tab:result_casme3}.  
Compared to recent state-of-the-art methods, our approach achieves significant improvements in both UF1 and UAR, demonstrating its notable advantage across multiple calssification task on larger datasets. Moreover, during the experiment, we observe that the 3-class task faces severe class imbalance, with the \textit{negative} class representing 65\% of the total samples and \textit{positive} less than 8\%. As a result, a considerable portion of samples were misclassified as \textit{negative}. Nevertheless, MER-CLIP scores a classification accuracy of 62\% on the \textit{positive} category, showing that the AU-prompts enable fine-grained learning of ME movements and strengthen robustness under extreme sample distribution conditions.

Additionally, to further validate our model’s performance, we conduct 7-class experiments on the recently released DFME dataset, which is the largest ME dataset currently available. The experiments are performed on Test A and Test B, which have different category distributions. The number of samples in each category of Test B tends to be balanced, while the category imbalance problem in Test A is relatively severe. Table \ref{tab:result_dfme} displays the detailed results. On Test A, MER-CLIP achieves a UF1 of 0.5024 and UAR of 0.5115, surpassing the best results from He et al. \cite{dfme2024ccac}, who achieved UF1 of 0.4123 and UAR of 0.4210. Similarly, on Test B, MER-CLIP obtains strong results, further underscoring its robustness across different splits. Notably, the methods proposed by He et al. and Wang et al. ranked first and second on the two test sets of the Dynamic Micro-expression Automatic Recognition Challenge on the fourth Chinese Conference on Affective Computing \cite{dfme2024ccac}, establishing them as recent state-of-the-art approaches for DFME.

The confusion matrices for the 7-class classification tasks on CAS(ME$)^3$, DFME Test A, and DFME Test B also reveal a common phenomenon: for \textit{negative} subclasses such as \textit{anger}, \textit{disgust}, \textit{fear}, and \textit{sadness}, the model frequently misclassifies and confuses these categories. We attribute this to the unique challenges of MEs: in contrast to MaEs, MEs are suppressed, so distinctive AUs commonly seen in negative MaEs (e.g., AU9, AU10, AU15, AU20) appear less frequently. Instead, they are often replaced by ambiguous, inhibitory AUs, such as AU14 and AU24 et al. Unlike the more distinct 3-class division (\textit{negative}, \textit{positive}, and \textit{surprise}), the AU combinations in different negative MEs overlap, resulting in less clear inter-class distinctions. Additionally, individual variability contributes to large intra-class differences. This frequent confusion in negative emotions leads to a tricky challenge in MER. Compared to other models, our approach, through AU-specific multimodal semantic alignment and refined emotional feature learning, captures the significant ME movement patterns and achieves substantial improvement in MER performance.
\subsection{Ablation Study}\label{sec_abla}

\begin{table*}[t]
\caption{Ablation results of significant proposed modules in MER-CLIP on three datasets.}
\centering
\label{tab:ablation_total}
\begin{threeparttable}
\begin{tabular}{c c c c c c c c c c}
  \toprule
  AU-Guided & Emotion & Progressive& \multirow{2}{*}{Data Augmentation} & \multicolumn{2}{c}{DFME TestA} & \multicolumn{2}{c}{DFME TestB} & \multicolumn{2}{c}{SAMM}\\
  \cmidrule(r){5-6}\cmidrule(r){7-8}\cmidrule(r){9-10}
  Cross-Modal Alignment & Inference & Training &  & UF1 & UAR & UF1 & UAR & UF1 & UAR\\
  
  \cmidrule(r){1-10}
  \XSolidBrush & \XSolidBrush & - & \XSolidBrush &0.3497 &0.3685 & 0.3219 & 0.3306 & 0.6500 & 0.6409 \\
  \XSolidBrush & \Checkmark & - & \XSolidBrush &0.3836 &0.3911 & 0.3230 & 0.3363 & 0.7196 & 0.7027 \\
  \Checkmark & \XSolidBrush & \XSolidBrush  & \XSolidBrush &0.4176 & 0.4209 & 0.3833 & 0.4103 & 0.7673 & 0.7460 \\
  
  \Checkmark & \Checkmark & \XSolidBrush & \XSolidBrush &0.4355 & 0.4458 & 0.4200 & 0.4257 & 0.8080 & 0.8209 \\
  \Checkmark & \Checkmark & \Checkmark & \XSolidBrush & 0.4639 & 0.4776 & 0.4563 & 0.4692 & 0.8043 & 0.8115 \\
  \Checkmark & \Checkmark & \Checkmark & \Checkmark & \textbf{0.5024} & \textbf{0.5115} & \textbf{0.5128} & \textbf{0.5120} & \textbf{0.8321} & \textbf{0.8434} \\
  Emo-Guided & - & - & \Checkmark &0.4222 &0.4247 & 0.4380 & 0.4326 & 0.7968 & 0.7938 \\
  
  \bottomrule
\end{tabular}
\begin{tablenotes}
\item[1] Emo-Guided: Emotion-guided text prompts (see Section \ref{sec_abla}).
\end{tablenotes}
\end{threeparttable}
\end{table*}

\begin{table*}[t]
\caption{Ablation results of data augmentation on UniformerV2 and MER-CLIP.}
\centering
\label{tab:ablation_aug}
\begin{threeparttable}
\begin{tabular}{c c c c c c c c}
  \toprule
  \multirow{2}{*}{Model} & Data & \multicolumn{2}{c}{DFME TestA} & \multicolumn{2}{c}{DFME TestB} & \multicolumn{2}{c}{SAMM}\\
  \cmidrule(r){3-4}\cmidrule(r){5-6}\cmidrule(r){7-8}
   & Augmentation& UF1 & UAR & UF1 & UAR & UF1 & UAR\\
  
  \cmidrule(r){1-8}
  \multirow{4}{*}{UniformerV2} & Base &0.3497 &0.3685 & 0.3219 & 0.3306 & 0.6500 & 0.6409 \\
   & + LSFM &0.4005 &0.4104 & 0.3655 & 0.3605  & 0.7040 & 0.6916 \\
   & + AugMix &0.3990 &0.4060 & 0.3894 & 0.3916 & 0.7182 & 0.7158 \\
   & + LSFM + AugMix &\textbf{0.4242} &\textbf{0.4446} & \textbf{0.4061} & \textbf{0.4121} & \textbf{0.7295} & \textbf{0.7046} \\
  \cmidrule(r){1-8}
  \multirow{4}{*}{MER-CLIP} & Base & 0.4639 & 0.4776 & 0.4563 & 0.4692 & 0.8043 & 0.8115 \\
   & + LSFM & 0.4747 & 0.4862 & 0.4645 & 0.4698 & 0.8041 & 0.8226 \\
   & + AugMix & 0.4618 & 0.4650 & 0.4859 & 0.4869 & 0.8204 & 0.8212 \\
   & + LSFM + AugMix & \textbf{0.5024} & \textbf{0.5115} & \textbf{0.5128} & \textbf{0.5120} & \textbf{0.8321} & \textbf{0.8434} \\
  
  \bottomrule
\end{tabular}
\begin{tablenotes}
\item[1] LSFM: LocalStaticFaceMix.
\end{tablenotes}
\end{threeparttable}
\end{table*}

\begin{table}[t]
\caption{Ablation Study on AU Description Design.}
\label{tab:au_design}
\centering
\begin{threeparttable}
\begin{tabular}{c@{\hspace{0.5em}}c@{\hspace{0.5em}}c@{\hspace{0.5em}}c@{\hspace{0.5em}}c@{\hspace{0.5em}}c@{\hspace{0.5em}}c@{\hspace{0.5em}}c}
  \toprule
  \multirow{2}{*}{Order} & \multirow{2}{*}{Prompt Type} & \multicolumn{2}{c}{DFME TestA} & \multicolumn{2}{c}{DFME TestB} & \multicolumn{2}{c}{SAMM} \\
  \cmidrule(r){3-4} \cmidrule(r){5-6} \cmidrule(r){7-8}
  & & UF1 & UAR & UF1 & UAR & UF1 & UAR \\
  \midrule
  F & Action-Oriented & \textbf{0.5024} & \textbf{0.5115} & \textbf{0.5128} & \textbf{0.5120} & \textbf{0.8321} & \textbf{0.8434} \\
  F & FACS-Based & 0.4942 & 0.5008 & 0.4720 & 0.4824 & 0.8032 & 0.7980 \\
  S & Action-Oriented & 0.4767 & 0.4861 & 0.4790 & 0.4664 & 0.8025 & 0.8289 \\
  \bottomrule
\end{tabular}
\begin{tablenotes}
\item[1] F: Fixed order of AU combination descriptions; S: Shuffled order.
\item[2] Action-Oriented: Refined descriptions focus on AU movements (\textit{raising the cheeks}); FACS-Based: Standard FACS terminology (\textit{Cheek Raiser}).
\end{tablenotes}
\end{threeparttable}
\end{table}

\begin{table}[t]
\caption{Ablation Study on Template Quantity.}
\centering
\label{tab:template_num}
\begin{threeparttable}
\begin{tabular}{c c c c c c c}
  \toprule
  Templates & \multicolumn{2}{c}{DFME TestA} & \multicolumn{2}{c}{DFME TestB} & \multicolumn{2}{c}{SAMM}\\
  \cmidrule(r){2-3}\cmidrule(r){4-5}\cmidrule(r){6-7}
  Number & UF1 & UAR & UF1 & UAR & UF1 & UAR\\
  
  \cmidrule(r){1-7}
  1 &0.4799 &0.4848 & 0.4937 & 0.5045 & 0.7932 & 0.8241  \\
  3 &0.4799 &0.4891 & 0.4950 & 0.4950 & 0.8247 & 0.8386  \\
  5 &0.4847 &0.4930 & 0.5004 & 0.5013 & 0.8207 & 0.8385  \\
  7 & \textbf{0.5024} & \textbf{0.5115} & \textbf{0.5128} & \textbf{0.5120} & \textbf{0.8321} & \textbf{0.8434} \\
  10 &0.4664 &0.4781 & 0.4628 & 0.4753 & 0.8175 & 0.8337  \\
  
  \bottomrule
\end{tabular}
\end{threeparttable}
\end{table}

To evaluate the contribution of each component in MER-CLIP and their collective impact on model performance, we conduct an extensive ablation study. The experiments are performed on the DFME dataset, which is selected for its larger sample size and relatively stable experimental conditions, with a 7-class MER setup. Additionally, we evaluate on the more challenging SAMM dataset, which features diverse ethnic representations and follows a 3-class MER task. The results, summarized in Tables \ref{tab:ablation_total}--\ref{tab:template_num}, examine the role of our proposed model components and various data augmentation techniques, while also investigating the effects of AU description design and the number of textual templates.

\subsubsection{Effect of Proposed Model Components}
Table \ref{tab:ablation_total} presents the ablation results for key components in MER-CLIP. The baseline model, which consists of a standard UniformerV2 encoder followed by a linear classification head without any specialized modules, achieves UF1 = 0.3497 and UAR = 0.3685 on DFME TestA, indicating its limited capacity to capture subtle ME movements. The introduction of the AU-Guided Cross-Modal Alignment significantly improves performance across all datasets, highlighting the effectiveness of AU-based vision-language alignment in facilitating fine-grained motion feature extraction. The addition of the Emotion Inference Module further enhances results, yielding improvements of approximately 2\%–7\% in UF1 and UAR. This suggests that the transformer-based emotion head effectively bridges the gap between CLIP’s motion-aligned features and emotion semantics, thereby improving classification robustness. However, when we remove the AU-Guided module while retaining only the Emotion Inference Module following the UniformerV2 encoder, performance drops sharply, only marginally outperforming the baseline model. This indicates that without the AU-Guided CLIP module, the transformer head functions merely as a classifier rather than an interpretable emotion-semantic mapper. Furthermore, the Progressive Training Strategy provides an additional performance boost, demonstrating its role in stabilizing training and feature adaptation.

Additionally, we compare MER-CLIP, which integrates the aforementioned modules, with a conventional Emo-guided CLIP model that uses generic emotion labels as textual prompts and performs cross-modality alignment via CLIP contrastive loss. As shown in Table \ref{tab:ablation_total}, AU-guided MER-CLIP significantly outperforms its Emo-guided counterpart. This result highlights that directly aligning ME visual features with generic emotion-based textual features is insufficient. Instead, leveraging anatomically grounded AU textual descriptions to capture subtle ME movements before learning emotion semantics proves to be a more effective approach for addressing the challenges of MER.
\subsubsection{Impact of Data Augmentation}
To assess the impact of data augmentation, we compare different augmentation strategies for both UniformerV2 and MER-CLIP, as shown in Table \ref{tab:ablation_aug}. LocalStaticFaceMix enhances the model’s robustness by introducing facial appearance variations while preserving essential motion cues, leading to a UF1 gain of approximately 3\% in MER-CLIP and up to 6\% in UniformerV2. AugMix alone also yields notable improvements, particularly on DFME TestB, indicating its effectiveness in enhancing generalization. The combination of LocalStaticFaceMix and AugMix achieves the best results across all datasets, demonstrating their complementary benefits. Notably, these augmentation techniques contribute substantial performance gains, underscoring the necessity and importance of effective data augmentation strategies in MER, where data scarcity is a significant challenge.
\subsubsection{Effect of AU Description Design}
Table \ref{tab:au_design} examines the effect of different AU description formulations on model performance. We analyze two key aspects: (1) whether AU descriptions are action-oriented or based on standard FACS terminology and (2) whether AU combinations follow a fixed (F) or shuffled (S) order. The results show that action-oriented descriptions consistently outperform FACS-based ones, suggesting that intuitive, motion-related phrases enhance cross-modal alignment with visual features. Additionally, maintaining a fixed order for AU combinations leads to better performance compared to a shuffled arrangement, indicating that a consistent semantic structure benefits model comprehension.
\subsubsection{Effect of Template Quantity}
Table \ref{tab:template_num} investigates the impact of varying the number of textual templates used in prompts. All templates employed in MER-CLIP are listed in Table \ref{tab_temlist}. Using a single template (Template No.1) results in suboptimal performance, while increasing the number to 7 (Template No.1-7) yields the best results, with UF1 reaching 0.5024 on DFME TestA and 0.5128 on TestB. However, further increasing the number of templates to 10 (Template No.1-10) degrades performance, likely because the introduction of excessive variability disrupts feature consistency—especially when the text encoder remains frozen.

\begin{table}[t]
    \caption{List of Templates Employed in This Study.}\label{tab_temlist}
    \centering
        \begin{tabular}{cc} 
        \toprule
        No. & Template\\
        \cmidrule(r){1-2} 
        1 & This micro-expression involves \{\}. \\
        2 & Key features of this micro-expression are \{\}. \\
        3 & This micro-expression is characterized by \{\}. \\
        4 & One can identify this micro-expression by \{\}. \\
       5 & This micro-expression typically manifests as \{\}. \\
        6 & The hallmark of this micro-expression is \{\}. \\
        7 & An identifiable trait of this micro-expression is \{\}. \\
        8 & Observing \{\} helps identify this micro-expression. \\
        9 & \makecell{The facial muscle movements in this \\ micro-expression include \{\}.} \\
        10 & This micro-expression can be recognized through \{\}. \\
        \bottomrule
        \end{tabular}
\end{table}
\subsection{Interpretability Analysis}

\begin{figure}[t] 
    \centering
    \includegraphics[scale=0.52]{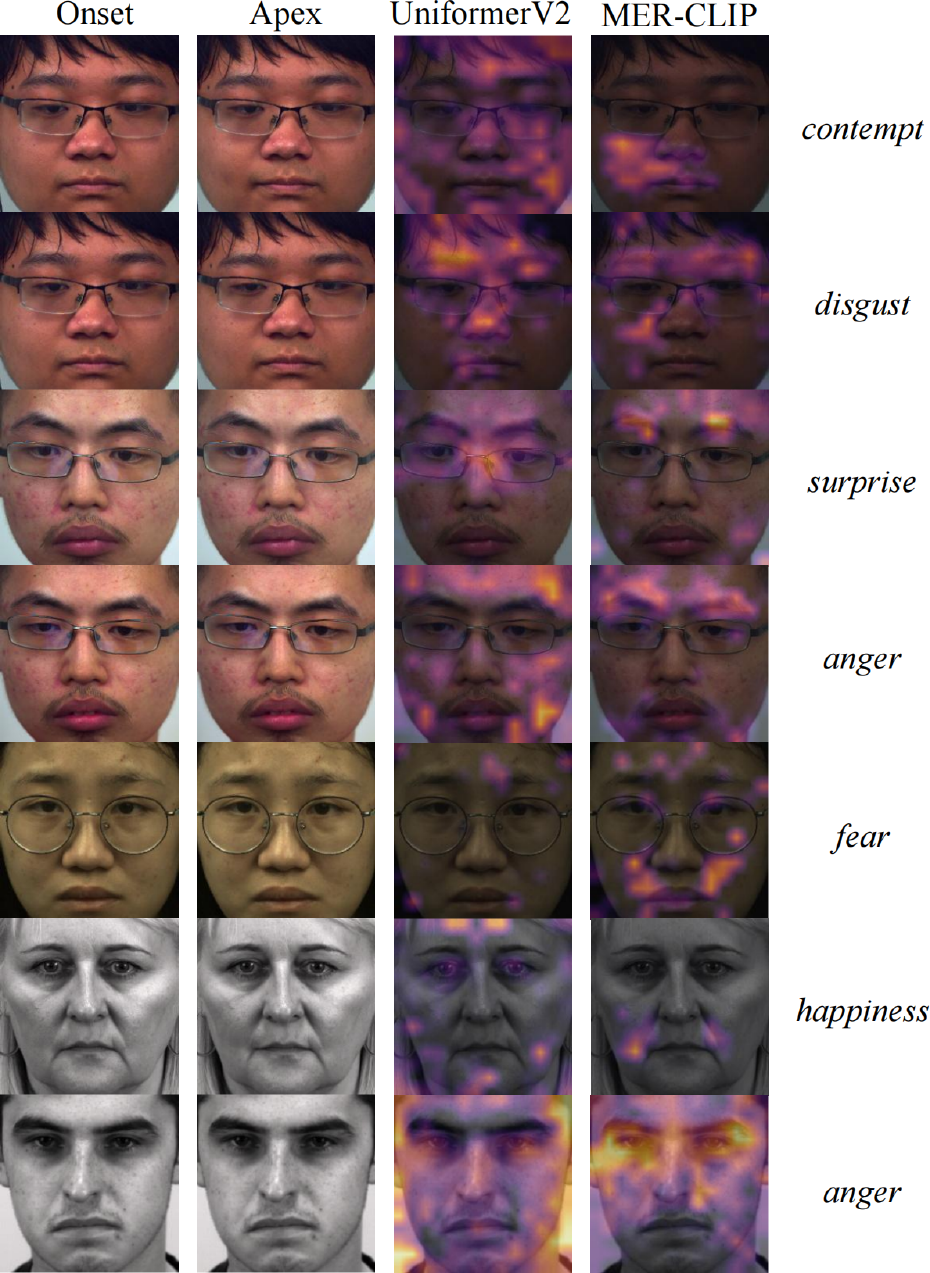}
    \caption{GradCam visualizations of several ME samples selected from DFME and SAMM to demonstrate how MER-CLIP captures subtle facial movements.}
    \label{fig_atten}
\end{figure}

To intuitively demonstrate the effectiveness of the MER-CLIP model, we conduct two kinds of visualization analyses. Specifically, we employ the Gradient-weighted Class Activation Mapping (GradCam) \cite{selvaraju2017grad} to generate instance-level discriminative visualization maps of MER-CLIP's vision encoder and the UniformerV2. Figure \ref{fig_atten} presents visualization results of ME samples from DFME and SAMM datasets, which are excluded from the training set. For clarity, we select the GradCam result of the apex frame (the frame with the most significant change) from each map sequence that best highlight the performance of capturing subtle dynamics. Additionally, we present both onset (the start of the expression) and apex frames in the raw sequences to better illustrate the MEs. 
From the visualizations, it is evident that the UniformerV2, which lacks specialized design for subtle facial dynamics, exhibits suboptimal localization capabilities. For instance, in the first-row example, it fails to focus on the critical left AU12 (Lip Corner Puller), instead dispersing attention across irrelevant facial regions. More critically, in the last-row case, the model erroneously assigns the highest attention to the background edges rather than the facial region. In contrast, our proposed MER-CLIP, which incorporates AU-Guided cross-modal alignment, effectively localizes ME-related AU activations. In the second row, MER-CLIP not only attends to AU4 (Brow Lowerer) but also accurately captures the activation of the left-side AU10 (Upper Lip Raiser), which is a more distinctive cue for disgust. Furthermore, in the penultimate row, MER-CLIP focuses solely on the mouth corner movements, disregarding irrelevant regions, thereby demonstrating its superior ability to filter out noise and enhance interpretability in ME recognition. These visual evidences confirm that explicit AU guidance enables targeted attention to transient facial action units, substantially enhancing the model's discriminative power for ME analysis.

\begin{figure}[t] 
    \centering
    \includegraphics[scale=0.25]{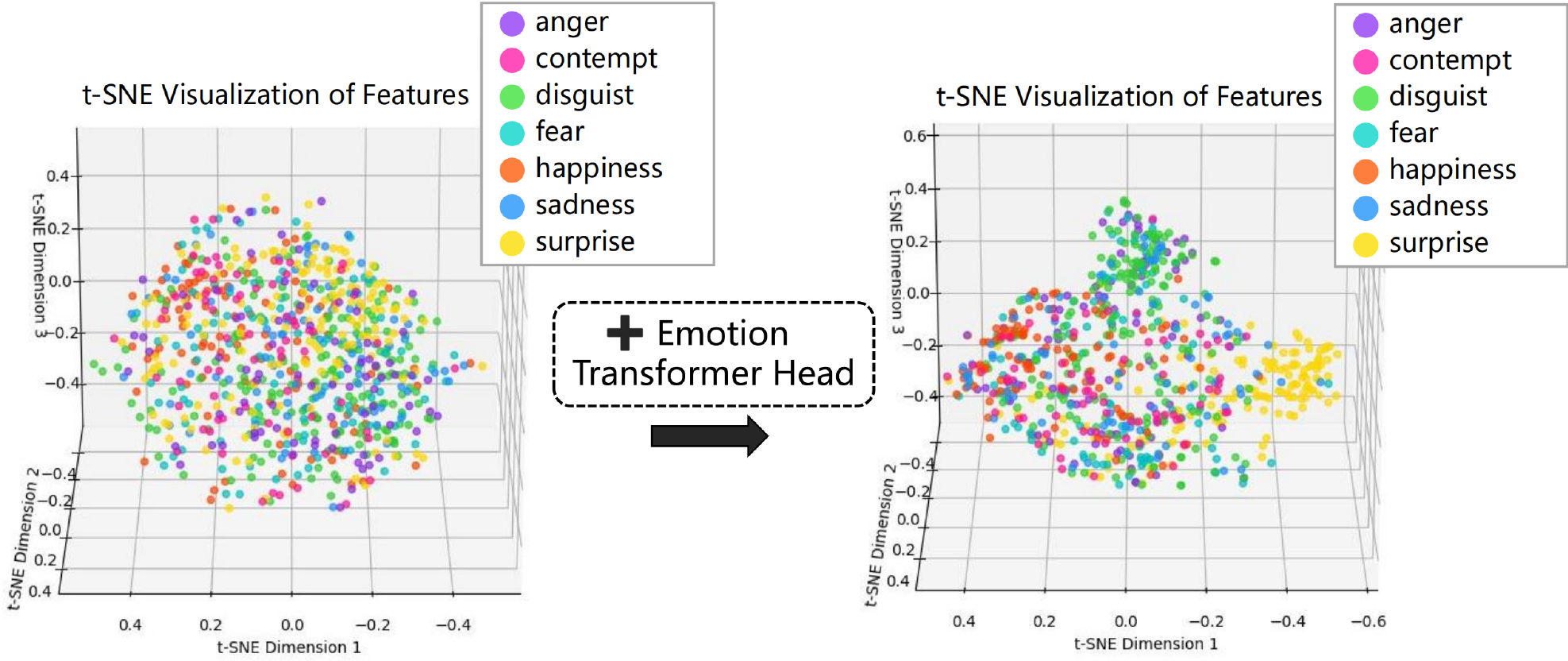}
    \caption{The t-SNE visualizations of ME features for classifying TestA and TestB in the DFME dataset before and after employing the emotion transformer head.}
    \label{fig_tsne}
\end{figure}

Additionally, the t-distributed stochastic neighbor embedding (t-SNE) \cite{van2008visualizing} is utilized to visualize the ME feature distributions for classifying TestA and TestB in the DFME dataset. As shown in Figure \ref{fig_tsne}, when using a simple linear classification head, the ME features for different emotion categories exhibit poor separability, leading to suboptimal classification performance. However, after introducing a lightweight transformer head for further emotion feature learning, intra-class clustering became more compact, and the decision boundaries between different categories became more distinct, particularly for \textit{disgust} and \textit{surprise}. Nonetheless, given the inherent difficulty of 7-class MER, there is still significant room for improvement in distinguishing other \textit{negative} emotions, consisting with the results shown in confusion matrices.

\section{Conclusion}\label{sec_conclusion}

In this paper, we introduced MER-CLIP as an initial exploration of applying CLIP to the MER task. By leveraging AU-guided textual prompts and cross-modal alignment, our approach enhances the learning of subtle ME movements and demonstrates the effectiveness of AU-based feature learning, offering a more reliable and interpretable alternative to direct emotion inference.

Despite its strong performance, MER-CLIP has limitations. It relies on accurate AU labels, which are labor-intensive to annotate, suggesting the potential for semi-supervised learning to reduce dependency on manual labeling. Additionally, the use of hand-crafted textual prompts introduces subjectivity, motivating future research on refining the text branch with learnable or fine-tuned prompts.  
Beyond model design, dataset challenges such as class imbalance and distinguishing similar negative emotions remain significant. Addressing these issues will be crucial for improving accuracy and robustness. 

Furthermore, current research largely follows Ekman’s discrete emotion theory, which may not fully capture emotional nuances. Future work could explore vision-language models for generating textual descriptions of ME dynamics, offering richer and more interpretable insights into facial expressions.

\bibliographystyle{IEEEbib}
\bibliography{ref}

\begin{IEEEbiography}[{\includegraphics[width=1in,height=1.25in,clip,keepaspectratio]{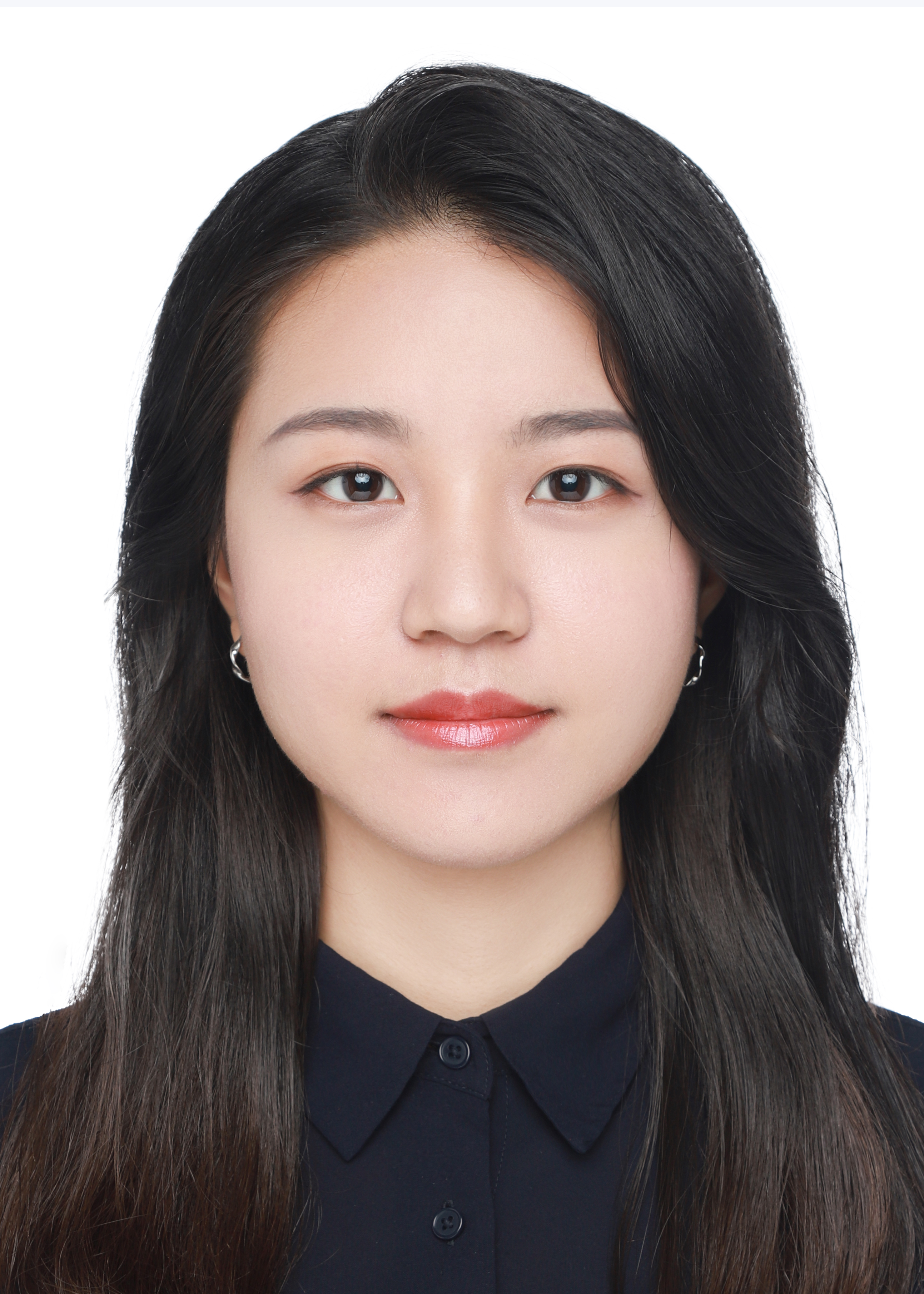}}]{Shifeng Liu} received the B.S degree in the School of Gifted Young from University of Science and Technology of China (USTC). She is currently working toward the PhD degree from School of Artificial Intelligence and Data Science, USTC. Her research interests include automatic micro-expressions analysis, affect computing, and human-computer interaction (HCI). She has published several papers in refereed
conferences and journals, including ACM Multimedia Conference, ICME, IEEE Transactions on Affective Computing, Neural Networks, etc.
\end{IEEEbiography}
\begin{IEEEbiography}[{\includegraphics[width=1in,height=1.25in,clip,keepaspectratio]{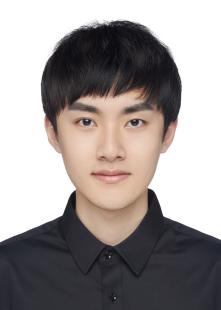}}]{Xinglong Mao} received the B.S degree in School of Data Science from University of Science and Technology of China (USTC), Hefei, China. He is currently working toward the PhD degree from School of Artificial Intelligence and Data Science, USTC. His research interests include automatic micro-expression analysis and affective computing. He has published several conference and journal papers in ACM Multimedia Conference, ICME, IEEE Transactions on Affective Computing, etc.
\end{IEEEbiography}
\begin{IEEEbiography}
[{\includegraphics[width=1in,height=1.25in,clip,keepaspectratio]{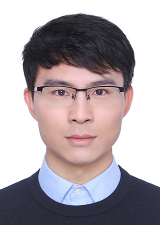}}]{Sirui Zhao} received the PhD
degree with the Department of Computer Science
and Technology from University of Science and
Technology of China (USTC).  He is also a faculty member with the USTC. His research interests
include automatic micro-expressions analysis,
human-computer interaction (HCI) and affect computing. He has published 30+ papers in refereed
conferences and journals, including ACM Multimedia Conference, KDD, ICME, IEEE Transactions on Affective Computing, ACM TOMM, Neural Networks, etc.
\end{IEEEbiography}
\begin{IEEEbiography}[{\includegraphics[width=1in,height=1.25in,clip,keepaspectratio]{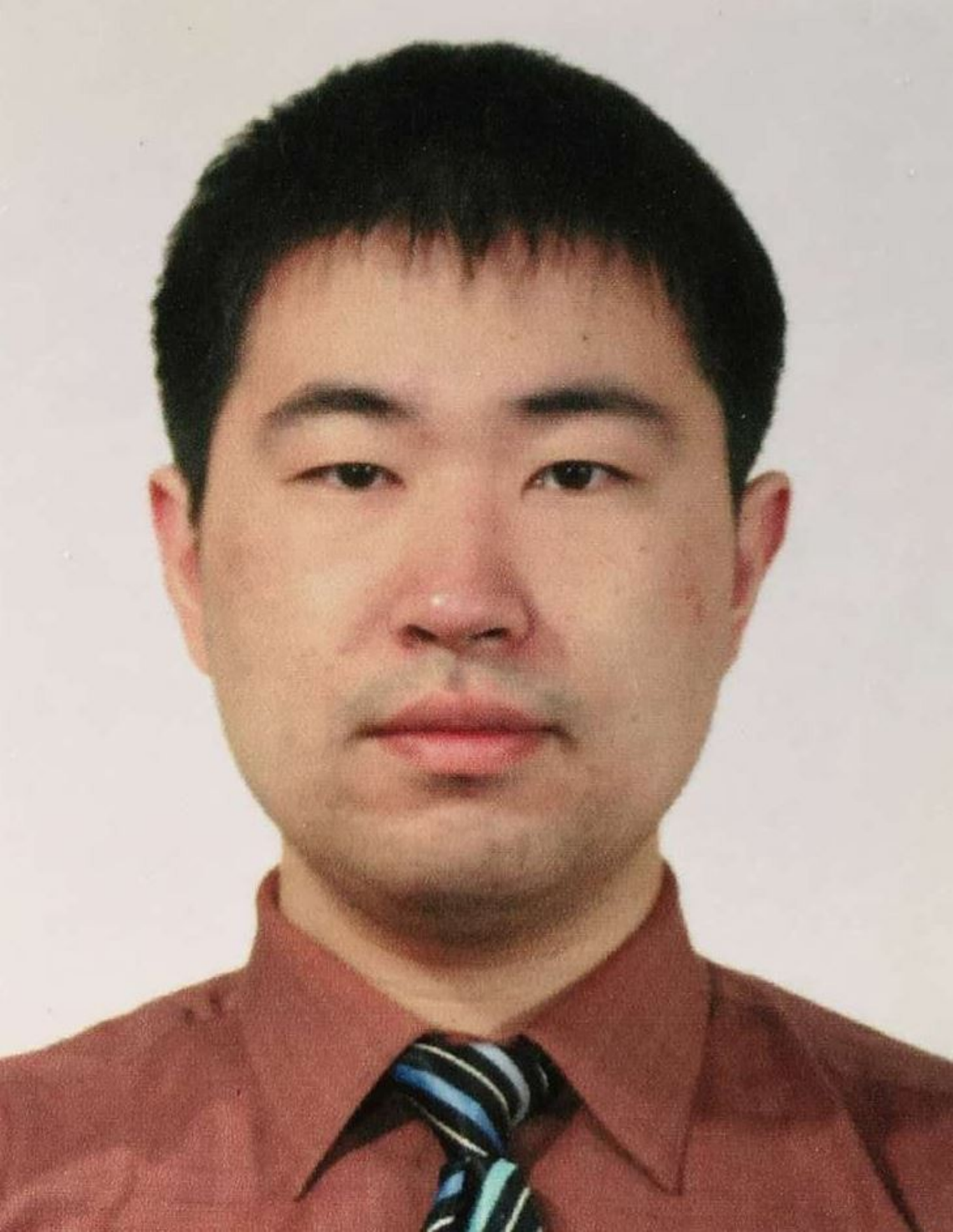}}]{Tong Xu}
received the Ph.D. degree in University of Science and Technology of China (USTC), Hefei, China, in 2016. He is currently working as a Professor of School of Computer Science and Technology, USTC. His research interests include Social Media Analysis, Multimodal Intelligence and other data mining-related techniques. He has authored 100+ journal and conference papers in the fields of social network and social media analysis, including IEEE TKDE, IEEE TMC, IEEE TMM, KDD, AAAI, ICDM, etc.
\end{IEEEbiography}

\begin{IEEEbiography}[{\includegraphics[width=1in,height=1.25in,clip,keepaspectratio]{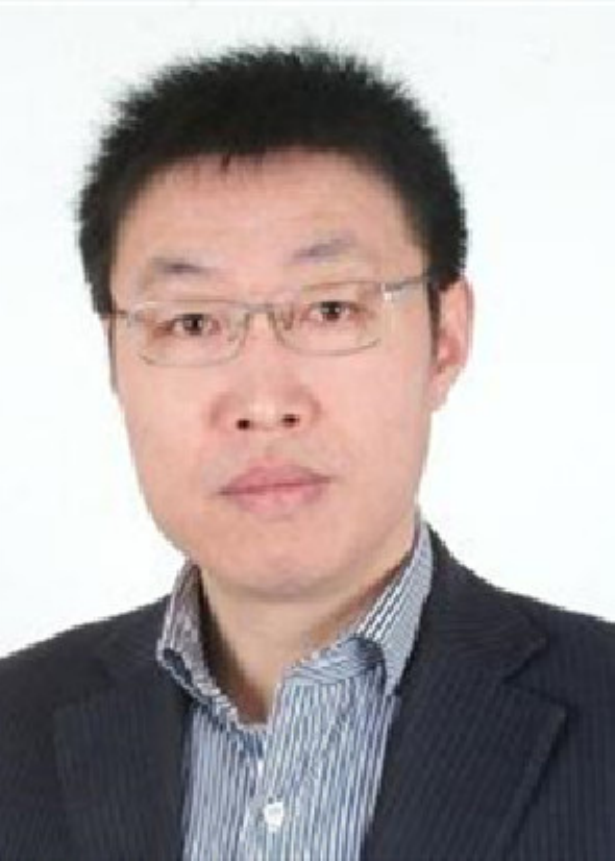}}]{Enhong Chen} (Fellow, IEEE) received the PhD degree from USTC. He is a professor of School of Computer Science and Technology, USTC.
His general area of research includes data mining
and machine learning, social network analysis, and
recommender systems. He has published more
than 200 papers in refereed conferences and journals, including IEEE Transactions on Knowledge
and Data Engineering, IEEE Transactions on Mobile
Computing, KDD, ICDM, NeurIPS, and CIKM. He
was on program committees of numerous conferences including KDD, ICDM, and SDM. His research is supported by the
National Science Foundation for Distinguished Young Scholars of China.
\end{IEEEbiography}

\end{document}